\documentclass[journal, 11pt,one column,twosides]{IEEEtran} 
\IEEEoverridecommandlockouts
\usepackage{graphicx}
\usepackage{caption}
\usepackage{subcaption}
\usepackage{color}
\usepackage{epsfig}
\usepackage{amssymb}
\usepackage{amsmath}
\usepackage{amsthm}
\usepackage{latexsym}
\usepackage{setspace}
\usepackage{bbm}
\usepackage{arydshln}
\usepackage{flushend}
\usepackage{textcomp}
\usepackage{amsfonts}
\usepackage{blindtext}
\usepackage{enumerate}
\usepackage{bbm}
\usepackage{upgreek}
\usepackage{hyperref}
\usepackage[bottom]{footmisc}
\usepackage{tikz}

\newcommand{\cA}{{\mathcal A}}
\newcommand{\cB}{{\mathcal B}}
\newcommand{\cC}{{\mathcal C}}

\newcommand{\cX}{{\mathcal X}}
\newcommand{\cY}{{\mathcal Y}}
\newcommand{\cZ}{{\mathcal Z}}

\usepackage{color}

\theoremstyle{remark}
\newtheorem*{remark*}{Remark}
\newtheorem*{remarks*}{Remarks}

\newcommand{\defn}{\triangleq}

\usepackage{secdot}
\allowdisplaybreaks
\sectiondot{subsection}
\sectiondot{subsubsection}
\usepackage{mleftright}
\usepackage{chngcntr}

\usepackage{breqn}

\begin{document}

\newcommand{\qeed}{\hfill $\blacksquare$}
\newcommand{\lnn}[1]{%
  \ln\left(#1\right)%
}
\newcommand\MC{{ \ - \!\!\circ\!\! - \ }}

\newcommand{\lnb}[1]{%
  \ln\mleft(#1\mright)%
}

\providecommand{\keywords}[1]{\textbf{\textit{Index terms---}} #1}
\theoremstyle{theorem}
\newtheorem{theorem}{Theorem}
\newtheorem{corollary}[theorem]{Corollary}
\newtheorem{lemma}[theorem]{Lemma}
\newtheorem{proposition}[theorem]{Proposition}
\theoremstyle{definition}
\newtheorem{definition}{Definition}
\title{Data Privacy for a $\rho$-Recoverable Function}
\author{Ajaykrishnan Nageswaran and Prakash Narayan$^\dag$ }
\maketitle
{\renewcommand{\thefootnote}{} \footnotetext{

\noindent 
$^\dag$A. Nageswaran and P. Narayan are with the Department of
Electrical and Computer Engineering and the Institute for Systems
Research, University of Maryland, College Park, MD 20742, USA.
E-mail: \{ajayk, prakash\}@umd.edu.\\ 

\noindent 
This work was supported by the U.S.
National Science Foundation under Grant CCF 1527354.
}}


\begin{abstract}
\label{abstract}
A user's data is represented by a finite-valued random variable. Given a function of the data, 
a querier is required to recover, with at least a prescribed probability, the value of the function
based on a query response provided by the user. The user devises the query response,
subject to the recoverability requirement, so as to maximize privacy of the data from the querier. 
Privacy is measured by the probability of error incurred by the 
querier in estimating the data from the query response. We analyze single and multiple independent 
query responses, with each response satisfying the recoverability requirement,
 that provide maximum privacy to the user. In the former setting, we also consider privacy for a 
predicate of the user's data.  Achievability schemes with explicit randomization mechanisms for 
query responses are given and their privacy compared with converse upper bounds. 
\end{abstract}



\section{Introduction}
\label{sec:intro}
Consider a (legitimate) user's data that is represented by a finite-valued random variable (rv)
with known probability mass function (pmf). 
A querier wishes to compute a given function of the data from a user-provided query response 
which is a suitably randomized version of the data. The user devises the query response so as 
to enable the querier to recover from it the function value with a prescribed accuracy while 
maximizing privacy of the data, i.e., minimizing the likelihood of the querier learning the data 
value from it. A generalization entails the user devising multiple independent such query responses 
with each query response adhering to the prescribed recoverability requirement, while maximizing 
overall privacy.

\par We consider a new and rudimentary formulation of this setting
in which the user forms a query response 
from which the querier can recover the function value with probability at least $\rho$, 
$0\leq\rho\leq 1$. Under this requirement, the chosen query response must afford maximum data privacy. 
Specifically, the query response must inflict -- on the querier's best estimate from it of 
the data value -- a maximum probability of error. Beginning with a single query response, 
we give an explicit characterization of a 
randomization mechanism that enables $\rho$-recoverability of the function value and yields 
the corresponding maximum privacy, termed $\rho$-privacy.
In particular, our query-response scheme is tantamount 
to an ``add-noise'' mechanism with the user computing first the function value and then adding 
to it a suitable value-dependent noise. Our optimal single query response depends on the
pmf of the data rv only in a limited way through associated minentropies. 
Furthermore, when privacy is sought for a predicate of the user data, we 
obtain a characterization of \textit{predicate $\rho$-privacy} and an explicit randomization mechanism that 
attains it. Next, when the querier elicits $n\geq 1$ $\rho$-recoverable and independent query responses, 
privacy of user data can degrade while accuracy of function estimation by the querier improves. 
We provide a converse upper bound for maximum privacy with respect to such responses, i.e., 
$\rho$-privacy, for every $n$. When $0.5<\rho\leq 1$, this upper bound decays exponentially in 
$n$ to a limit which is the querier's data-estimation error on the basis of a knowledge of the exact 
function value (i.e., corresponding to $\rho=1$). The rate of this decay is shown to be 
(the Kullback-Leibler divergence) $D\left(\text{Ber($0.5$)}||\text{Ber($\rho$)}\right)$. 
We provide an explicit add-noise achievability scheme with privacy that converges to the mentioned 
limit at the same exponential rate. When $0\leq\rho\leq 0.5$, we again provide an explicit add-noise 
achievability scheme. While it remains unknown whether the corresponding privacy is optimal, this scheme 
is shown to prevent the querier from estimating exactly the function value for \textit{any} $n$. 
Neither achievability scheme depends on a knowledge of the pmf of the data rv. Finally, these two 
achievability schemes are shown to be asymptotically superior in privacy to 
a scheme made up of (conditionally) independent and identically distributed (i.i.d.) repetitions 
of our optimal single query response; this is done by means of suitable asymptotic approximations 
of privacy in terms of Chernoff radii. The superiority of the former schemes is enabled by 
rendering an estimation by the querier of the exact function value to be more 
error-prone than by the latter scheme, while conforming to the $\rho$-recoverability requirement.

\par An explanation of our approach is in order. In a model for private function computation,
the querier can possess initial knowledge or beliefs of the user's data in the form of a family
of {\em prior} pmfs that describe said data. Accordingly, the user must fashion a query response 
or responses that assure data privacy in the form of an adequate querier's
probability of data-estimation error for {\em every} prior in said family. As indicated in
Section~\ref{sec:discussion}, the minmax of the probability of data-estimation error
(maximum and minimum, respectively, over query responses and prior
pmfs) serves as a minimum guarantee of privacy for user data. 
In this approach, our concept of ${\rho}$-privacy developed below plays a basal role
whose operational significance is clear also if the querier's uncertainty regarding the
user's data were reflected by a (single) known pmf or if the user's data were known to be
{\em generated} by said pmf.
It should be added that the maximum probability of error criterion is
eminently tractable -- as our work shows -- 
compared with more discerning measures, e.g., $L_1$- or $L_2$-distances between 
user data and the querier's estimate of it. The latter measures serve to penalize deviation of
the querier's estimate of user data from its true value, a discriminating feature missing in our
work (and one which is currently under study).

\par Our approach is in the spirit of prior works that deal with information leakage of a 
user's private data with associated nonprivate correlated  data. A randomized version of the 
nonprivate data is released publicly under a constraint on the expected distortion between the 
nonprivate and public data. For instance, in~\cite{Mondero10, CalmonFawaz12, MakhdoumiFawaz13}, 
leakage as measured by the mutual information between the private and public data is minimized with respect 
to the ``channel'' from the former to the latter, while constraining a distortion between the 
nonprivate and public data. In a more elaborate setting~\cite{Sankar13}, temporally i.i.d. 
private and nonprivate data that are correlated across multiple 
users are encoded into a bin index. With this index and additional side-information as inputs, a 
decoder reconstructs the nonprivate data under a distortion constraint. Privacy is 
gauged by the conditional entropy rate of the private data given the decoder's inputs, and 
achievable privacy-distortion pairs are characterized. These works are based on principles of 
rate distortion theory. A variant model in~\cite{Liao18} considers private and public data as the input and 
output of a channel with a ``hard'' distortion requirement between them being met with probability 
$1$. Based on a concept of ``maximal leakage'' introduced in~\cite{Issa16}, 
privacy-recoverability tradeoffs are characterized with privacy measured in terms of $\alpha$-R{\'e}nyi 
divergence. In a separate vein, in~\cite{Huang17} a possibly randomized function of the private and 
nonprivate data is released publicly while constraining the expected distortion between the nonprivate 
and public data. Measuring privacy in terms of a minimal expected loss function of private data and its estimate 
based on public data, optimal privacy mechanisms are ``learned'' in binary and Gaussian settings using 
techniques based on generative adversarial nets~\cite{Goodfellow14}. By comparison, for finite-valued 
data and query responses, upon limiting ourselves to information leakage as a probability of error and 
recoverability as a (pointwise) conditional probability of error, we obtain exact and approximate 
utility-privacy tradeoffs for single and multiple query responses, respectively. This is in contrast 
to a prior approach~\cite{Asoodeh18} in which maximum a posteriori (MAP) estimates of 
private and nonprivate data are formed on basis of a randomized version of the latter which is made public. 
The private, nonprivate and public data are assumed to form a Markov chain. Under a constraint on the 
probability of estimating correctly the private data, mechanisms are sought for said randomization so 
as to maximize correct MAP estimation of the nonprivate data.

\par An important movement that has received dominant attention in recent years is differential privacy, 
introduced in~\cite{Dwork06, DworkSmith06} and explored further in~\cite{McSherry07,Blum08, Bassily13,Kasi14}, 
among others. Consider a database that hosts multiple users' data that, in our framework, 
constitutes a data vector. The notion of differential privacy stipulates that altering
a data vector slightly leads only to a near-imperceptible change in the corresponding
probability distribution of the output of the privacy mechanism, i.e., 
query responses that are randomized 
functions of data vectors. We note that unlike in differential
privacy, our work lacks a notion of closeness of datasets.
Upon imposing a differential privacy constraint, there exists a large 
body of work that seeks to maximize function recoverability by minimizing a discrepancy cost between function 
value and randomized query response; a sampling is mentioned below. \textit{In contrast, our work maximizes privacy 
under a constraint on recoverability}, and may be viewed as a companion approach. Considering a 
class of linear functions of data, tradeoffs between recoverability as measured by the worst-case 
$L_2$-distance (over user data) 
between function value and query response, and differential privacy, are examined in~\cite{Hardt10}. 
Similar tradeoffs for add-noise differential private mechanisms with an additional restriction 
are characterized in~\cite{Geng16}. Other pertinent works include parameter estimation~\cite{Smith08}, 
empirical-frequency-of-data estimation~\cite{Bassily15} and distribution estimation~\cite{Duchi16, Kairouz16, Ye17}, 
all under differential privacy constraints. A relaxation of the concept of differential privacy is 
examined in~\cite{Bassily13} in the form of distributional differential privacy as part of a 
larger framework of ``coupled-worlds privacy." Distributional differential privacy requires 
the mentioned indistinguishability to hold for a random data vector over probability distributions in a specified family
(to which our allusion above to the querier's initial knowledge of a family of prior pmfs for the 
user's data is redolent). This is in contrast to a worst-case requirement over the 
family of all probability distributions of the data vector in a differential privacy context. 

\par Directions other than differential privacy have also been pursued. 
As mentioned in~\cite{Wasserman10}, these include 
studies based on clustering (e.g.,~\cite{Sweeney02}), $t$-closeness (e.g.,~\cite{Li07}), 
data pertubation (e.g.,~\cite{Evfi02}), etc; see~\cite{Wasserman10} for a comprehensive list. Other 
methods include $(\rho_1,\rho_2)$-privacy (e.g.,~\cite{Evfi03}), confidence intervals (e.g.,~\cite{Agrawal00}), 
and cryptographic approaches (e.g.,~\cite{Blum05}).

\par Our model for $\rho$-recoverable function computation with associated privacy is 
described in Section~\ref{sec:prelim}. The $\rho$-privacy and predicate $\rho$-privacy for 
a single query response are characterized in Section~\ref{sec:rhopriv}, and $\rho$-privacy is extended to 
multiple independent query responses 
in Section~\ref{sec:rhoprivn}. The inadequacy of (conditionally) i.i.d. repetitions of the optimum 
$\rho$-privacy scheme of 
Section~\ref{sec:rhopriv} in the context of Section~\ref{sec:rhoprivn} is brought out in 
Section~\ref{sec:inadequacy-Wo}. The concluding Section~\ref{sec:discussion} mentions
unanswered questions even in our simple setting of multiple independent query responses.



\section{Preliminaries}
\label{sec:prelim}

A (legitimate) user's data is represented by a rv $X$ taking values in a finite 
set $\cX$ with $|\cX|=r,$ say, and of known pmf $P_X$ with $P_X\left(x\right)>0,$ $x\in\cX.$ 
Throughout, we shall consider a given mapping $f:\cX\rightarrow\cZ=\{0,1,\ldots,k-1\},$ $2\leq k\leq r.$ 
For a realization $X=x$ in $\cX,$ a querier -- who does not know $x$ -- wishes to compute $f\left(x\right)$ 
from a \textit{query response} (QR) $F\left(x\right)$ provided by the user, where $F\left(x\right)$ is a rv 
with values in $\cZ.$ A QR must satisfy the following recoverability condition. 

\vspace{0.20cm}

\begin{definition}
\label{def:rho-recov}
Given $0\leq\rho\leq 1,$ a QR $F\left(X\right)$ is 
$\rho$\textit{-recoverable} if 
\begin{equation}
\label{eq:recov1}
P\left(F\left(X\right)=f\left(x\right)\big|X=x\right)\geq 
\rho,\hspace{4mm}x\in\cX.
\end{equation} 
Condition~\eqref{eq:recov1} can be written equivalently 
in terms of a stochastic matrix $W:\cX\rightarrow\cZ$ with the 
requirement 
\begin{equation}
\label{eq:recov2}
W\big(f\left(x\right)|x\big)\geq \rho,\hspace{4mm}x\in\cX;
\end{equation}

\noindent which, too, will constitute a $\rho$\textit{-recoverable} QR.
Such a $\rho$-recoverable $F\left(X\right)$ or $W$ will be termed $\rho$-QR. Note that $\rho$-recoverability 
in~\eqref{eq:recov1},~\eqref{eq:recov2} does not depend on $P_X$.
\end{definition}
\vspace{0.2cm}

\begin{definition}
\label{def:addnoise}
A $\rho$-QR $F(X)$ will be called an \textit{add-noise} $\rho$-QR if it can be expressed as
\begin{equation}
\label{eq:addnoise1}
F(X)=f(X)+N \text{ mod $k$} 
\end{equation}

\noindent where $N$ is a $\cZ$-valued rv that satisfies
\begin{equation}
 \label{eq:mark_cond}
 N\MC f(X)\MC X
\end{equation}

\noindent and with conditional pmf given by
\begin{align}
P\left(N=i\big|X=x\right)
&=P\left(N=i\big|f(X)=f(x),X=x\right)\nonumber\\
&=P\left(N =i\big|f (X)=f(x)\right)\label{eq:addnoise2}\\
&=V\left(i+f(x)\text{ mod $k$ }\big|f(x)\right)\label{eq:addnoise3}
\end{align}

\noindent for some stochastic matrix $V:\cZ\rightarrow\cZ$ with $V(i|i)\geq\rho, \ i \in \cZ$;
we shall refer to it also as add-noise $\rho$-QR $V$.
Thus, an add-noise $\rho$-QR is obtained by adding to the function value 
$f(x)$ a noise $N$ whose (conditional) pmf can depend on $f(x)$.
\end{definition}
\vspace{0.2cm}

\noindent By~\eqref{eq:addnoise1}-\eqref{eq:addnoise3}, an add-noise $\rho$-QR $F(X)$ with $V:\cZ\rightarrow\cZ$ 
has the following property:
\begin{equation}
\label{eq:addnoise4}
P\left(F\left(X\right)=i\big|f(X)=f(x),X=x\right)=V\left(i|f(x)\right),
\hspace{4mm}i \in \cZ,\hspace{2mm}x\in\cX.
\end{equation}

\vspace{0.2cm}

\begin{definition}
\label{def:priv}
Denoting by $Z$ the rv $F\left(X\right)$ with values in $\cZ,$ the 
\textit{privacy} of a $\rho$-QR $F\left(X\right)$ (or 
equivalently $\rho$-QR $W$) satisfying~\eqref{eq:recov1} 
$\big($respectively~\eqref{eq:recov2}$\big)$ is
\begin{equation}
\label{eq:priv} 
\pi_\rho\left(F\right)= 
\pi_\rho\left(W\right)=\min_g\hspace{1.5mm}
P\big(g\left(Z\right)\neq X\big)
\end{equation}

\noindent where the minimum is over all estimators $g:\cZ\rightarrow\cX$ 
of $X$ on the basis of $F(X).$ Clearly, the minimum in~\eqref{eq:priv} is attained 
by the MAP estimator $g_{MAP}=g_{MAP(W)}:\cZ\rightarrow\cX$ given by
\begin{equation}
\label{eq:map1}
g_{MAP(W)}(i)=\arg\hspace{0.1cm}\max_{x\in\cX}\hspace{1.5mm}
P_X\left(x\right)W\left(i|x\right),\hspace{4mm}i\in\cZ
\end{equation}

\noindent so that~\eqref{eq:priv} equals 
$P\big(g_{MAP(W)}\left(Z\right)\neq X\big).$ When $F(X)$ is an add-noise 
$\rho$-QR $V$ as in Definition~\ref{def:addnoise},
we shall denote $\pi_\rho(F)$ in~\eqref{eq:priv} by $\pi_\rho(V).$ The corresponding 
minimum in~\eqref{eq:priv} will be denoted by $P\big(g_{MAP(V)}\left(Z\right)\neq X\big)$ where 
\begin{equation}
\label{eq:map2}
g_{MAP(V)}(i)=\arg\hspace{0.1cm}\max_{x\in\cX}\hspace{1.5mm}P_X\left(x\right)
V\big(i|f(x)\big),\hspace{4mm}i\in\cZ.
\end{equation}

\noindent Ties in~\eqref{eq:map1} and~\eqref{eq:map2} are broken arbitrarily.
\end{definition}
\vspace{0.2cm}
\noindent \textit{Remark}: We assume throughout that the querier knows $P_X$ and $W$ for computing the MAP estimate 
in~\eqref{eq:map1}.
\begin{definition}
\label{def:rhopriv}
For each $0\leq\rho\leq 1,$ the maximum privacy that can be attained by a $\rho$-QR is termed 
$\rho$\textit{-privacy} and denoted by $\pi\left(\rho\right),$ i.e.,
\begin{equation}
\label{eq:rhopriv1}
\pi\left(\rho\right)= \max_{\substack{W : W \left( f \left(x\right) | x 
\right)\geq\rho \\ x\in\cX}}  \pi_\rho\left(W\right).
\end{equation}
\end{definition}

\noindent\textit{Remark}: That the maximum in~\eqref{eq:rhopriv1} exists will be seen below.
\vspace{0.2cm}
\par The following simple lemma shows when a $\rho$-QR $W$ is also an
add-noise $\rho$-QR, and will be helpful in our proofs of achievability of privacy by $\rho$-QRs.

\begin{lemma}
\label{lem:addnoise}
Given $0\leq\rho\leq 1$, for a $\rho$-QR $W: \cX\rightarrow\cZ$ with identical rows for all 
$x\in f^{-1}(i), \ i \in \cZ$, there exists an add-noise $\rho$-QR $V=V(W): \cZ \rightarrow \cZ$ 
with the same privacy, i.e., with ${\pi}_{\rho} (V) = {\pi}_{\rho} (W)$. Conversely, for an add-noise 
$\rho$-QR $V: \cZ \rightarrow \cZ$, there exists a $\rho$-QR $W = W (V): \cX \rightarrow \cZ$ with 
identical rows as above, and with ${\pi}_{\rho} (W) = {\pi}_{\rho} (V)$.
\end{lemma}

\noindent\textit{Proof}: For a stochastic matrix $W: \cX \rightarrow \cZ$
which satisfies~\eqref{eq:recov2} and has rows 
$\left\{\left(W(i'|x),\text{ }i' \in \cZ \right), x \in \cX\right\}$
that are identical for all $x\in f^{-1}(i), \ i \in \cZ$, consider
a stochastic matrix $V=V(W):\cZ\rightarrow\cZ$ given by
\begin{equation}
\label{eq:lem-addnoise1}
V(i|j)=W(i|x)\text{ for every $x\in f^{-1}(j), \ i, j \in \cZ$}
\end{equation}

\noindent and an associated add-noise QR $F'(X)$ defined as in~\eqref{eq:addnoise1}-\eqref{eq:addnoise3} 
with $V$ as above. Since $V (i | i) \geq \rho, \ i \in \cZ$, in~\eqref{eq:lem-addnoise1}, $F'(X)$ 
is an add-noise $\rho$-QR. To see that ${\pi}_{\rho} (V) = {\pi}_{\rho} (W)$, we have
\begin{align}
\label{eq:lem-addnoise2}
P\Big(g_{MAP(V)}\big(F'\left(X\right)\big)= X\Big) \ = \ 
\sum_{i \in \cZ} \max_{x\in\cX} P \big( X = x, F' (X) = i\big)
\end{align}

\noindent where in the right-side,
\begin{eqnarray*}
P \big( X = x, F' (X) = i \big) &=& \sum_{j \in \cZ} P \big( X = x, f (X) = j, F' (X) = i \big)  \\
&=&   P \big( X = x, f (X) = f(x) \big) P_{F' \left( X \right) | f \left( X \right)} 
\left(i | f(x)\right), \ \ {\rm by \ \eqref{eq:addnoise4}}  \\
&=& P_X (x) V \big( i | f\left(x\right)\big)  \\
&=& P_X (x) W \big( i | x \big), \ \ {\rm by \ \eqref{eq:lem-addnoise1}}. 
\end{eqnarray*}

\noindent Hence, by~\eqref{eq:lem-addnoise2},
\begin{eqnarray*}
1 - \pi_{\rho} (V) &=& P\Big(g_{MAP(V)}\big(F'\left(X\right)\big)= X\Big)  \\
&=& \sum_{i \in \cZ} \max_{x\in\cX} P_X (x) W \big( i | x \big)  \\
&=& 1 - \pi_{\rho} (W). 
\end{eqnarray*}

Conversely, given an add-noise $\rho$-QR $V: \cZ\rightarrow\cZ$, 
consider a stochastic matrix $W=W(V): \cX\rightarrow\cZ$ 
with identical rows for all $x\in f^{-1}(i), \ i \in \cZ$,
defined by~\eqref{eq:lem-addnoise1}. By the same steps as above, 
this $W$ is a $\rho$-QR and, furthermore, ${\pi}_{\rho} (W) = {\pi}_{\rho} (V)$. 
\qeed
\vspace{0.2cm}

\par A justification of our model above is warranted. Our choice of the probability of error 
as a measure of recoverability as well as privacy is driven by considerations of 
tractability and obtaining exact answers, as indicated in Section~\ref{sec:intro}. In 
particular, it enables us to identify optimal or asymptotically optimal $\rho$-QRs in our 
achievability proofs. In symmetry with the pointwise measure of recoverability in $X=x$ 
in~\eqref{eq:recov1} or~\eqref{eq:recov2}, it would be preferable to consider a redolent measure 
of privacy in~\eqref{eq:priv},~\eqref{eq:rhopriv1} that is pointwise in $Z=i$, viz.
\[ 
\max_{\substack{W : W \left( f \left(x\right) | x 
\right)\geq\rho \\ x\in\cX}}\hspace{0.2cm}\max_{i\in\cZ} \ P\big(g_{MAP(W)}\left(Z\right)\neq X \ \big | \ Z=i\big).
\]

\noindent However, such conservatism leads to intractability; by contrast, our liberal choice 
in~\eqref{eq:priv},~\eqref{eq:rhopriv1}, which is equivalent to
\begin{equation}
 \label{eq:max_measure}
\max_{\substack{W : W \left( f \left(x\right) | x 
\right)\geq\rho \\ x\in\cX}}\hspace{0.2cm}\max_{i\in\cZ} \ P\big(g_{MAP(W)}\left(Z\right)\neq X, \ Z=i\big),
\end{equation}

\noindent makes for comprehensive analysis as will be seen below; the equivalence of~\eqref{eq:max_measure} 
is explained in the remark following the proof of Theorem~\ref{thm:rhopriv} in the next section.

\par Concerning $\rho$-recoverability, we add that there is no loss of generality 
in~\eqref{eq:recov1},~\eqref{eq:recov2} by not considering an \textit{estimate} of $f(X)$ 
on the basis of $F(X)$; this is so, because the user can emulate any such estimation strategy 
of the querier to produce yet another $\rho$-QR.

\par Lastly, a seemingly more general setting comprising ``private data $X$, correlated nonprivate 
data $Y$ and publicly released data $Z$'' is addressed below; see Remark (iii) after the proof of 
Proposition~\ref{prop:predicate}.
\vspace{0.2cm}



\section{$\rho$-Privacy for a Single Query Response}
\label{sec:rhopriv}
A characterization of $\rho$-privacy is provided by obtaining first an upper bound 
for $\pi(\rho)$ and then identifying explicitly an add-noise $\rho$-QR whose 
privacy meets the bound.
\par Let 
\begin{equation}
\label{eq:prob}
x^*=\arg\hspace{0.1cm}\max_{x\in\cX}\hspace{1.5mm}P_X\left(x\right),\hspace{
8mm} x^*_i=\arg\hspace{0.1cm}\max_{x\in f^{-1}\left(i\right)}\hspace{1.5mm}P_X\left(x\right),\hspace{4mm}i\in\cZ
\end{equation}

\noindent and suppose that $x^*\in f^{-1}\left(i^*\right)$ for some $i^*\in\cZ,$ 
where $x^*,i^*$ and $x_i^*,$ $i\in\cZ,$ need not be unique. Further, 
set 
\begin{equation}
\label{eq:rhoc}
\rho_c=\frac{P_X\left(x^*\right)}{\sum\limits_{i\in\cZ}P_X\left(x_i^*\right)}
\end{equation}
and observe that $1/k\leq\rho_c <1,$ where the left inequality is by
\[\frac{P_X\left(x^*\right)}{\sum\limits_{i\in\cZ}P_X\left(x_i^*\right)}
\geq\frac{P_X\left(x^*\right)}{\sum\limits_{i\in\cZ}P_X\left(x^*\right)}=\frac
{1}{k}.\]

\noindent The following choice of $\rho$-QR $W=W_o:\cX\rightarrow\cZ$ 
will play a material role in the achievability proof of $\rho$-privacy in 
Theorem~\ref{thm:rhopriv} below:
\begin{equation}
\label{eq:optchannel1}
W_o\left(i|x\right)=
\begin{cases}
\max\{\rho_c,\rho\},&\hspace{4mm}i=f(x)\\
\Big(1-\max\{\rho_c,\rho\}\Big)\frac{P_X\left(x_i^*\right)}{\sum\limits_{
l\neq f(x)}P_X\left(x_l^*\right)},&\hspace{4mm}i\neq f(x), \ \ x \in \cX, \ i \in \cZ.
\end{cases}
\end{equation} 

\noindent We note that $W_o$ has the property that for each $i\in\cZ$, 
all rows of $W_o$ corresponding to $x\in f^{-1}\left(i\right)$ are identical. 
By dint of Lemma~\ref{lem:addnoise}, the associated stochastic 
matrix $V_o:\cZ\rightarrow\cZ$ given by
\begin{align}
\label{eq:optchannel2}
V_o\left(i|j\right)&=W_o\left(i|x\right)\hspace{2mm}\text{for every $x\in 
f^{-1}\left(j\right)$}\nonumber\\
&=\begin{cases}
\max\{\rho_c,\rho\},&\hspace{4mm}i=j\\
\Big(1-\max\{\rho_c,\rho\}\Big)\frac{P_X\left(x_i^*\right)}{\sum\limits_{
l\neq j}P_X\left(x_l^*\right)},&\hspace{4mm}i\neq j,\ \ i,j \in \cZ
\end{cases}
\end{align}

\noindent will be also of consequence in achieving $\rho$-privacy.  
\vspace{0.2cm}
\par An exact characterization of $\rho$-privacy is provided by

\begin{theorem}
\label{thm:rhopriv}
$\rho$-privacy equals
\begin{align}
\label{eq:rhopriv}
\pi\left(\rho\right)=1-\text{\normalfont{max}}\left\{P_X\left(x^*\right),
\rho\sum_{i \in \cZ} P_X\left(x_i^*\right)\right\}=1-\text{\normalfont{max}}
\left\{\rho_c,\rho\right\}\sum_{i \in \cZ} 
P_X\left(x_i^*\right),\hspace{4mm} 0\leq\rho\leq 1.
\end{align}

\noindent Furthermore, $\rho$-privacy is achieved by the $\rho$-QR $W_o$ 
in~\eqref{eq:optchannel1} 
and, additionally, by the add-noise $\rho$-QR $V_o$ in~\eqref{eq:optchannel2}.
\end{theorem}

\vspace{0.2cm}

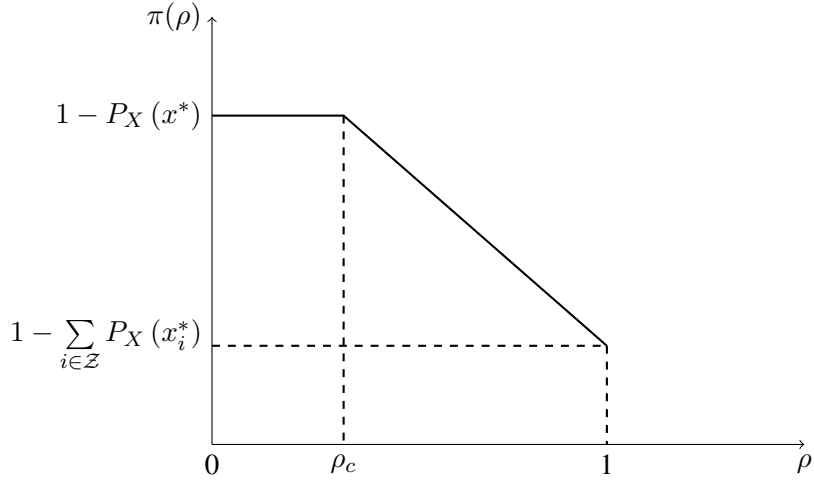
\begin{figure}
\centering
\begin{tikzpicture}[scale=0.875]
\draw[->] (0,0) -- (9,0) node[anchor=north] {$\rho$};
\draw	(0,0) node[anchor=north] {0}
		(2,0) node[anchor=north] {$\rho_c$}
		(6,0) node[anchor=north] {1};
\draw	(0,1.5) node[anchor=east]{$1-\sum\limits_{i\in\cZ}P_X\left(x_i^*\right)$}
		(0,5) node[anchor=east] {$1-P_X\left(x^*\right)$};
\draw[->] (0,0) -- (0,6.5) node[anchor=east] {$\pi(\rho)$};

\draw[thick] (0,5) -- (2,5) -- (6,1.5);
\draw[thick,dashed] (2,5) -- (2,0);
\draw[thick,dashed] (6,1.5) -- (6,0);
\draw[thick,dashed] (0,1.5) -- (6,1.5);
\end{tikzpicture}
\caption{$\pi(\rho)$ vs. $\rho.$}
\label{fig:privacy_graph}
\end{figure}

\noindent
\textit{Remarks}:
\begin{enumerate}[(i)]

\item The choice of $W_o$ and $V_o$ in~\eqref{eq:optchannel1} and~\eqref{eq:optchannel2}, and
the value of $\rho$-privacy in~\eqref{eq:rhopriv}, depend on $P_X$ only
through $P_X\left(x_i^*\right), i \in \cZ$, i.e., $P_X\left(f^{-1}(i)\right)2^{-H_{min}(P_i)}$, 
$i\in\cZ$, where $H_{min}(P_i)$ is the minentropy of the pmf $P_i=\left(P_X(x)/P_X\left(f^{-1}(i)\right), 
\ x\in f^{-1}(i)\right)$.

\item By Theorem~\ref{thm:rhopriv},
\[\pi(\rho)=
\begin{cases}
1-P_X\left(x^*\right),&0\leq\rho\leq\rho_c\\
1-\rho\sum\limits_{i\in\cZ}P_X\left(x_i^*\right),&\rho_c\leq\rho\leq 1
\end{cases}\]

and is plotted in Fig.~\ref{fig:privacy_graph}. In particular, for $0\leq\rho\leq\rho_c,$ 
$\pi(\rho)=1-P_X\left(x^*\right)$ and is the error of a MAP estimator of $X$ 
without any observation. For $\rho=1$, $\pi(1)=1-\sum\limits_{i\in\cZ}P_X\left(x_i^*\right)$ 
is the error of a MAP estimator of $X$ on the basis of $f(X).$

\item The $\rho$-privacy achieving $\rho$-QR $W_o$ and the 
corresponding  add-noise $\rho$-QR $V_o$ in 
Theorem~\ref{thm:rhopriv} are not unique. For instance, see Remark (ii) after the 
proof of Proposition~\ref{prop:predicate}, Remark (ii) following Theorem~\ref{thm:rhoprivn-achiev1} 
and the first part of the Remark following Theorem~\ref{thm:rhoprivn-achiev2}.
\end{enumerate}

\vspace{0.2cm}

\noindent\textit{Proof}: That the two characterizations of $\pi(\rho)$ 
in~\eqref{eq:rhopriv} are identical follows by straightforward 
manipulation. We first show that $\rho$-privacy cannot exceed the 
right-side(s) of~\eqref{eq:rhopriv}, and then identify a $\rho$-QR that attains it.
\vspace{0.2cm}\\
\textit{Converse:} Clearly 
\[
P\big(g_{MAP(W)}\left(Z\right)=X\big)\geq P_X\left(x^*\right) 
\]

\noindent and for every $W:\cX\rightarrow\cZ$ satisfying~\eqref{eq:recov2},
\[
P\big(g_{MAP(W)}\left(Z\right)=X\big)=\sum_{i \in \cZ} \max_{
x\in\cX}\hspace{1.5mm}P_X\left(x\right)W\left(i|x\right)
\geq\sum_{i \in \cZ} \max_{x\in 
f^{-1}\left(i\right)}P_X\left(x\right)W\left(i|x\right)\geq 
\rho\sum_{i \in \cZ} P_X\left(x^*_i\right)
\]

\noindent leading to
\begin{align}
\label{eq:thm-rhopriv-pf1}
P\big(g_{MAP(W)}\left(Z\right)= X\big)\geq 
\max\left\{P_X\left(x^*\right),\rho\sum_{i \in \cZ} P_X\left(x_i^*\right)\right\}.
\end{align}

\noindent Hence
\begin{equation}
\label{eq:thm-rhopriv-pf2}
\pi_{\rho}\left(W\right)= P\big(g_{MAP(W)}\left(Z\right)\neq X\big)\leq 
1-\max\left\{P_X\left(x^*\right),\rho\sum_{i \in \cZ}
P_X\left(x_i^*\right)\right\},\hspace{4mm}0\leq\rho\leq 1
\end{equation}

\noindent so that the same upper bound, valid for all $W: \cX \rightarrow \cZ$ 
subject to \eqref{eq:recov2}, applies to $\pi(\rho)$, too. 
\vspace{0.2cm}\\
\noindent\textit{Achievability}: We show that the choice of the $\rho$-QR 
$W_o:\cX\rightarrow\cZ$ in~\eqref{eq:optchannel1} has privacy 
$\pi_\rho\left(W_o\right)$ equal to the right-side(s) of \eqref{eq:rhopriv}.
To this end,
\begin{align}
1-\pi_\rho\left(W_o\right)&=P\big(g_{MAP(W_o)}\left(Z\right)= X\big)\nonumber\\
&=\sum_{i \in \cZ} \max_{x\in\cX}\hspace{1.5mm}
P_X\left(x\right)W_o\left(i|x\right)\nonumber\\
&=\sum_{i \in \cZ} \max\left\{\max_{x\in 
f^{-1}(i)}\hspace{1.5mm}P_X\left(x\right)W_o\left(i|x\right),
\max_{x\not\in 
f^{-1}(i)}\hspace{1.5mm}P_X\left(x\right)W_o\left(i|x\right)\right\}
\nonumber\\
&=\sum_{i \in \cZ} \max\left\{P_X\left(x_i^*\right)\max\{
\rho_c,\rho\},\max_{x\not\in 
f^{-1}(i)}\hspace{1.5mm}P_X\left(x\right)W_o\left(i|x\right)\right\}
,\hspace{4mm}\text{by~\eqref{eq:optchannel1}}.\label{eq:thm-rhopriv-pf4}
\end{align}

\noindent We claim that
\begin{equation}
\label{eq:thm-rhopriv-pf5}
P_X\left(x_i^*\right)\max\{\rho_c,\rho\}\geq 
\max_{x\not\in 
f^{-1}(i)}\hspace{1.5mm}P_X\left(x\right)W_o\left(i|x\right), \ \ i \in \cZ
\end{equation}

\noindent whereupon~\eqref{eq:thm-rhopriv-pf4} becomes 
\begin{align*}
1-\pi_\rho\left(W_o\right)&=\max\{\rho_c,\rho\}\sum_{i \in \cZ} P_X\left(x_i^*\right)
\end{align*}

\noindent so that the privacy $\pi_\rho\left(W_o\right)$ equals the right-side(s) 
of~\eqref{eq:rhopriv}. It remains to establish~\eqref{eq:thm-rhopriv-pf5}. Considering 
first the case $0\leq\rho\leq\rho_c,$ we must show for each $x\not\in f^{-1}(i)$ that 
\begin{align*}
P_X\left(x_i^*\right)\rho_c &\geq P_X\left(x\right)W_o\left(i|x\right)\\
&=P_X(x)\left(1-\rho_c\right)\frac{P_X\left(x_i^*\right)}{\sum\limits_{j\neq 
f(x)}P_X\big(x_j^*\big)},\hspace{4mm}\text{by~\eqref{eq:optchannel1}}
\end{align*}

\noindent i.e.,
\begin{equation}
\label{eq:thm-rhopriv-pf6}
\frac{\rho_c}{1-\rho_c}\geq\frac{P_X\left(x\right)}{\sum\limits_{j\neq 
f(x)}P_X\big(x_j^*\big)}
\end{equation}

\noindent which, in turn, would follow if
\[\frac{\rho_c}{1-\rho_c}\geq\frac{P_X\left(x^*\right)}{\sum\limits_{j\neq 
f(x)}P_X\big(x_j^*\big)},\]

\noindent which is tantamount to showing that
\begin{equation}
\label{eq:thm-rhopriv-pf7}
\frac{\sum\limits_{j\neq i^*}P_X\big(x_j^*\big)}{\sum\limits_{j\neq 
f(x)}P_X\big(x_j^*\big)}\leq 1.
\end{equation}

\noindent Clearly,~\eqref{eq:thm-rhopriv-pf7} holds for each $x\not\in f^{-1}(i),$ as 
the denominator is either larger than or equal to the numerator 
for all $i \in \cZ$. For the case $\rho_c\leq\rho< 1,$ we must 
show~\eqref{eq:thm-rhopriv-pf6} with $\rho_c$ replaced by $\rho;$ this follows readily since
\begin{equation*}
\frac{\rho}{1-\rho}\geq\frac{\rho_c}{1-\rho_c},\hspace{4mm}\text{ 
$\rho_c\leq\rho<1$}.
\end{equation*}

\noindent For $\rho=1,$ we have by~\eqref{eq:optchannel1} that 
$W_o\left(i|x\right)=\mathbbm{1}\big(i=f\left(x\right)\big), \ x \in 
\cX,$ $ i \in \cZ$, whereby~\eqref{eq:thm-rhopriv-pf5} holds trivially. 
\par Finally, that the add-noise $\rho$-QR $V_o$ achieves $\rho$-privacy follows by Lemma 1.
\qeed
\vspace{0.3cm}\\
\noindent \textit{Remark}: The equivalence in~\eqref{eq:max_measure} is justified by the proof of Theorem~\ref{thm:rhopriv} 
which shows, in effect, that a common maximizer in
\begin{multline*}
\arg
\hspace{0.1cm}\max_{\substack{W : W \left( f \left(x\right) | x 
\right)\geq\rho \\ x\in\cX}}\hspace{0.2cm}\ P\big(g_{MAP(W)}\left(Z\right)\neq X\big)= \arg\hspace{0.1cm}
\max_{\substack{W : W \left( f \left(x\right) | x 
\right)\geq\rho \\ x\in\cX}}\hspace{0.2cm}\sum\limits_{i\in\cZ} \ P\big(g_{MAP(W)}\left(Z\right)\neq X, \ Z=i\big)\\=
\arg\hspace{0.1cm}\max_{\substack{W : W \left( f \left(x\right) | x 
\right)\geq\rho \\ x\in\cX}}\hspace{0.2cm}\max_{i\in\cZ} \ P\big(g_{MAP(W)}\left(Z\right)\neq X, \ Z=i\big)
\end{multline*}
\noindent is $W_o$ in~\eqref{eq:optchannel1}.

\vspace{0.3cm}

We close this section by considering $\rho$-privacy for a predicate $Y=h(X)$ of $X$, where 
$h:\cX\rightarrow\cY=\{0,1,\ldots,m-1\}$, $2\leq m\leq r$, is a given mapping. Analogously as in 
Definition~\ref{def:priv}, \textit{predicate privacy}\footnote{Notation used in the context of 
$\rho$-privacy will be primed throughout for predicate $\rho$-privacy.} for a $\rho$-QR $F(X)$ or $W$ 
in~\eqref{eq:recov1},~\eqref{eq:recov2} is

\[
\pi'_\rho\left(F\right)= 
\pi'_\rho\left(W\right)=P\big(g'_{MAP(W)}\left(Z\right)\neq Y\big)
\]

\noindent and \textit{predicate $\rho$-privacy} is

\begin{equation}
 \label{eq:predicate_priv}
 \pi'(\rho)= \max_{\substack{W : W \left( f \left(x\right) | x 
\right)\geq\rho \\ x\in\cX}}  \pi'_\rho(W).
\end{equation}

\noindent Clearly, $\pi'(\rho)\leq\pi(\rho)$, $0\leq\rho\leq 1$, and 
when $h:\cX\rightarrow\cY=\cX$ is the identity mapping, predicate $\rho$-privacy 
in~\eqref{eq:predicate_priv} specializes to $\rho$-privacy in~\eqref{eq:rhopriv1}.\\
\par Proposition~\ref{prop:predicate} below provides an exact characterization of $\pi'(\rho)$. 
Its proof builds on that for $\pi(\rho)$ in Theorem~\ref{thm:rhopriv}. The following additional 
notation is convenient. Define
\begin{gather*}
 j^* = \arg \max_{j\in\cY}\hspace{1.5mm}
P_X\left(h^{-1}(j)\right),\\
P_X(i,j) = P_X\left(f^{-1}(i)\cap h^{-1}(j)\right),\hspace{4mm} i\in\cZ,\hspace{4mm} j\in\cY,\\
j_i^* = \arg\max_{j\in\cY}\hspace{1.5mm}P_X(i,j),\hspace{4mm} i\in\cZ
\end{gather*}

\noindent and
\begin{equation}
 \label{eq:predicate_rhoc}
 \rho'_c = \frac{P_X\left(h^{-1}(j^*)\right)}{\sum\limits_{i\in\cZ}P_X\left(i,j_i^*\right)},
\end{equation}

\noindent where the maxima above need not be attained uniquely. Observe that 
$\max\left\{\frac{1}{m},\frac{1}{k}\right\}\leq\rho'_c \leq 1$. The right inequality is by
\[
\rho'_c = \frac{\sum\limits_{i\in\cZ}P_X\left(i,j^*\right)}
{\sum\limits_{i\in\cZ}P_X\left(i,j_i^*\right)}\leq 1
\]

\noindent while the left inequality follows from
\[
\frac{P_X\left(h^{-1}(j^*)\right)}{\sum\limits_{i\in\cZ}P_X\left(i,j_i^*\right)}\geq
\frac{P_X\left(h^{-1}(j^*)\right)}{\sum\limits_{i\in\cZ}P_X\left(f^{-1}(i)\right)}=
P_X\left(h^{-1}(j^*)\right)\geq\frac{1}{m}
\]

\noindent and
\[
\frac{P_X\left(h^{-1}(j^*)\right)}{\sum\limits_{i\in\cZ}P_X\left(i,j_i^*\right)}\geq
\frac{P_X\left(h^{-1}(j^*)\right)}{\sum\limits_{i\in\cZ}P_X\left(h^{-1}(j_i^*)\right)}\geq 
\frac{P_X\left(h^{-1}(j^*)\right)}{\sum\limits_{i\in\cZ}P_X\left(h^{-1}(j^*)\right)}=\frac{1}{k}.
\]

\begin{proposition}
\label{prop:predicate}
Predicate $\rho$-privacy equals 
\begin{equation}
\pi'(\rho) =1-\max\left\{P_X\left(h^{-1}(j^*)\right),\rho\sum\limits_{i\in\cZ}
P_X\left(i,j_i^*\right)\right\} 
=1-\max\left\{\rho'_c,\rho\right\}\sum\limits_{i\in\cZ}P_X\left(i,j_i^*\right), 
\hspace{4mm} 0\leq\rho\leq 1. 
\label{eq:predicate_rho_priv}
\end{equation}
\end{proposition}
\vspace{0.2cm}
\noindent \textit{Proof}: Starting with the converse part, we have that 
\[P\big(g'_{MAP(W)}\left(Z\right)= Y\big) \geq P_X\left(h^{-1}(j^*)\right)  \]
 
\noindent and for every $\rho$-QR $W:\cX\rightarrow\cZ$ in~\eqref{eq:recov2},
\[
 P\big(g'_{MAP(W)}\left(Z\right)= Y\big) = \sum\limits_{i\in\cZ}\max_{j\in\cY} 
 \sum\limits_{x\in h^{-1}(j)} P_X(x) W(i|x)
 \geq \rho \sum\limits_{i\in\cZ}\hspace{0.1cm}\max_{j\in\cY} P_X(i,j)=\rho \sum\limits_{i\in\cZ}P_X(i,j_i^*)
\]

\noindent where the first equality uses
\[P\left(Y=j\big | X=x,Z=i\right) = \mathbbm{1} \left(j=h(x)\right).\]

\noindent The converse proof is completed similarly as for Theorem~\ref{thm:rhopriv}.

\par Turning to the achievability part, consider the $\rho$-QR $W_o':\cX\rightarrow\cZ$ specified for 
$x\in\cX$, $i\in\cZ$, $j\in\cY$, as follows.\\
For $\rho'_c=1$
\begin{equation}
\label{eq:predicate_channel_trivial}
W'_o(i|x)=\mathbbm{1}\left(f(x)=i\right)
\end{equation}

\noindent and for $\rho'_c<1$,
\begin{equation}
\label{eq:predicate_channel}
W'_o(i|x)=
\begin{cases}
\max\{\rho'_c,\rho\}+\left(1-\max\{\rho'_c,\rho\}\right)
\frac{P_X\left(i,j_i^*\right)-P_X(i,j)}{\sum\limits_{l\in\cZ}
P_X\left(l,j_l^*\right)-P_X\left(h^{-1}(j)\right)},
&i=f(x), \ \ j=h(x)\\
\left(1-\max\{\rho'_c,\rho\}\right)
\frac{P_X\left(i,j_i^*\right)-P_X(i,j)}{\sum\limits_{l\in\cZ}
P_X\left(l,j_l^*\right)-P_X\left(h^{-1}(j)\right)},
& i\neq f(x), \ \ j=h(x).
\end{cases}
\end{equation} 

\noindent Since $\rho'_c<1$, observe in~\eqref{eq:predicate_channel} that
\[ \sum\limits_{l\in\cZ}P_X\left(l,j_l^*\right)-P_X\left(h^{-1}(j)\right) \geq  
\sum\limits_{l\in\cZ}P_X\left(l,j_l^*\right)-P_X\left(h^{-1}(j^*)\right) >0. \]

\noindent We show in Appendix~\ref{app:predicate_achiev} that $\pi'_{\rho}(W_o')$ equals 
the right-side of~\eqref{eq:predicate_rho_priv}.

\par This completes the proof of the proposition.
\qeed
\vspace{0.2cm}\\
\textit{Remarks}:
\begin{enumerate}[(i)]
\item Note that $W'_o$ has the property that for each $i$ in $\cZ$, $j$ in $\cY$, 
all rows corresponding to $x$ in $f^{-1}(i)\cap h^{-1}(j)$ are identical. Pursuant to 
Lemma~\ref{lem:addnoise}, $W'_o$ can be interpreted as an add-noise $\rho$-QR obtained by adding 
to $f(X)$ a noise $N$ that satisfies the Markov condition $N\MC f(X),h(X) \MC X$.

\item When $h$ is the identity mapping, $W'_o$ in~\eqref{eq:predicate_channel} \textit{does not} coincide 
with $W_o$ in~\eqref{eq:optchannel1}; in fact, $W'_o$ reduces to
\begin{equation*}
W''_o(i|x)=
\begin{cases}
\max\{\rho_c,\rho\}+\left(1-\max\{\rho_c,\rho\}\right)
\frac{P_X\left(x_i^*\right)-P_X(x)}{\sum\limits_{l\in\cZ}P_X\left(x_l^*\right)-P_X (x)},
&i=f(x),\\
\left(1-\max\{\rho_c,\rho\}\right)
\frac{P_X\left(x_i^*\right)-P_X(x)}{\sum\limits_{l\in\cZ}P_X\left(x_l^*\right)-P_X (x)},
& i\neq f(x).
\end{cases}
\end{equation*} 

\noindent By Proposition~\ref{prop:predicate} and Theorem~\ref{thm:rhopriv}, $\pi'_{\rho}\left(W''_o\right)
=\pi(\rho)=\pi_{\rho}\left(W_o\right)$. On the other hand, and unlike $W_o$, the $\rho$-QR $W''_o$ is 
not of the add-noise type in the sense of Definition~\ref{def:addnoise} and also depends on the entirety of $P_X$.

\item Proposition~\ref{prop:predicate} covers the setting when privacy is sought for a randomized function 
$Y$ of the data $X$ with the (finite-valued) rvs $X,Y$ having a given joint pmf. Specifically $\rho$-privacy 
for $Y$ corresponds to predicate $\rho$-privacy in Proposition~\ref{prop:predicate} with 
$\bar{X}$, $h(\bar{X})$, $f(\bar{X})$ and $F(\bar{X})$, where 
\[
\bar{X} = (X,Y),\ \ h(\bar{X}) = Y,\ \ f(\bar{X}) = f(X) ,\ \ F(\bar{X})=Z,
\]
\noindent (with an abuse of notation in $f$).

\item In a similar vein, Proposition~\ref{prop:predicate} also covers the setting with ``private data $X$, correlated nonprivate 
data $Y$ and publicly released data $Z$'' alluded to in Section~\ref{sec:intro}. Formally, $\rho$-privacy 
for the mentioned setting equals predicate $\rho$-privacy in Proposition~\ref{prop:predicate} applied to 
$\tilde{X}$, $h(\tilde{X})$, $f(\tilde{X})$ and $F(\tilde{X})$, where
\[
\tilde{X} = (X,Y), \ \ h(\tilde{X}) = X, \ \ f(\tilde{X}) = Y ,\ \ F(\tilde{X})=Z.
\]

\end{enumerate}



\section{Multiple Independent Query Responses}
\label{sec:rhoprivn}

In a general setting, given a mapping $f:\cX\rightarrow\cZ,$ a 
querier wishes to compute $f(x),$ $x\in\cX,$ from $\rho$-QRs 
$\left\{\left(F_t(x),x\in\cX\right)\right\}_{t=1}^n,$ $n\geq 1.$ The rvs $\{F_t\left(X\right)\}_{t=1}^{n}$ 
are taken to be conditionally mutually independent, conditioned on $X,$ but not necessarily identically 
distributed, with \textit{each} $F_t\left(X\right)$ satisfying the $\rho$-recoverability 
condition~\eqref{eq:recov1}. Correspondingly, consider stochastic matrices 
$\left\{W_t:\cX\rightarrow\cZ\right\}_{t=1}^{n}$ such that
\begin{align}
P\big(F_1\left(X\right)=i_1,\ldots,F_n\left(X\right)=i_n\big|X=x\big)&=\prod_{
t=1}^n P\big(F_t\left(X\right)=i_t|X=x\big)\nonumber\\
&=\prod_{t=1}^n W_t\left(i_t|x\right),\hspace{4mm}x\in\cX,\hspace{2mm}\text{$i_1,\ldots,i_n$ 
$\in\cZ$}\label{eq:n_recov} 
\end{align} 

\noindent say, with \textit{each} $W_t$ satisfying~\eqref{eq:recov2}. Similarly, for 
add-noise $\rho$-QRs $F_t(X)$ as in 
Definition $2$ with $\{V_t:\cZ\rightarrow\cZ \text{ where }V_t(i|i)\geq\rho,$ $i\in\cZ\}_{t=1}^{n}$,
\begin{equation}
\label{eq:addnoisen} 
P\big(F_1\left(X\right)=i_1,\ldots,F_n\left(X\right)=i_n\big|X=x\big)=\prod_{
t=1}^n V_t\big(i_t|f(x)\big),\hspace{4mm}x\in\cX,\hspace{2mm}\text{$i_1,\ldots,i_n$ $\in\cZ.$}
\end{equation} 

\noindent In all contexts, denote $Z_t=F_t\left(X\right),$ $t=1,\ldots,n.$ 
\vspace{0.3cm}
\par  This formulation is apposite when the main objective of the querier is to improve its estimation 
accuracy (beyond $\rho$) of a \textit{given} $f(X)$ by soliciting multiple $\rho$-QRs whereas the 
user designs said $\rho$-QRs so as to maximize the privacy of its data $X$. A different formulation 
in which an ``adversarial querier'' elicits multiple $\rho$-QRs for various choices of functions in 
order to destroy data privacy by isolating the value of $X$, is beyond the scope of this paper.
\vspace{0.2cm}\\
\textit{Remark}: In addition to possibly eroding privacy,
multiple $\rho$-QRs enable the querier to estimate $f(X)$ with a probability 
that can exceed $\rho$. Precisely, for a MAP estimator $h_{\text{MAP}}$ of $f(X)$ on the basis of 
$\{F_t\left(X\right)\}_{t=1}^{n}$ in~\eqref{eq:n_recov}, we have 
\begin{equation}
\label{eq:func_estimate_lb}
 P\big(h_{\text{MAP}}\left(F_1(X),\ldots,F_n(X)\right)= f(X)\big) \geq 
 \max\left\{\rho,\max_{i\in\cZ}\hspace{1mm}P\left(f(X)=i\right),
 P\left(\text{\normalfont{Bin}}(n,\rho)\geq\left\lfloor\frac{n}{2}\right\rfloor+1\right)\right\}
\end{equation}

\noindent where Bin($n,\rho$) is a binomial rv with parameters $n\geq 1$ and $0\leq \rho\leq 1$. In 
particular, for $0.5<\rho\leq 1$, the right-side of~\eqref{eq:func_estimate_lb} tends to $1$ as $n\rightarrow\infty$. 
See Appendix~\ref{app:funct_estimate_lb} and Lemma~\ref{lem:binom-bds}.

\begin{definition}
For each $0\leq\rho\leq 1$ and $n\geq 1,$ the $\rho$-privacy that can be attained by 
$\rho$-QRs $\{F_t(X)\}_{t=1}^{n}$ 
as in~\eqref{eq:n_recov} with each $F_t(X)$ satisfying~\eqref{eq:recov1} 
$\big($or equivalently each $W_t$ satisfying~\eqref{eq:recov2}$\big)$ is

\begin{equation*}
\label{eq:n_priv1}  
 \pi_n\left(\rho\right)= \max_{\substack{W_1,\ldots,W_n :\\ W_t \left( f 
\left(x\right) | x \right)\geq\rho,\hspace{1mm} x\in\cX}}
\pi_{\rho}\left(W_1,\ldots,W_n\right), 
\end{equation*}

\noindent where
\begin{equation*}
\label{eq:n_priv2}
\pi_{\rho}\left(W_1,\ldots,W_n\right)=\min_{g_n}\hspace{1.5mm}
P\big(g_n\left(Z_1,\ldots,Z_n\right)\neq X\big),
\end{equation*}

\noindent with the minimum being taken over all estimators $g_n:\cZ^n\rightarrow\cX$ 
on the basis of $\left\{F_t(X)\right\}_{t=1}^n.$ Thus,
\begin{equation}
\label{eq:n-privacy-W}
\pi_{\rho}\left(W_1,\ldots,W_n\right)=P\big(g_{\text{MAP 
$\left(W_1,\ldots,W_n\right)$}}\left(Z_1,\ldots,Z_n\right)\neq X\big)
\end{equation}

\noindent where
\[g_{MAP\left(W_1,\ldots,W_n\right)}\left(i_1,\ldots,i_n\right)=\arg\hspace{0.1cm} 
\max_{x\in\cX}\hspace{1.5mm}
P_X\left(x\right)\prod\limits_{t=1}^{n}W_t\left(i_t|x\right),\hspace{4mm} 
i_1,\ldots,i_n\in\cZ.\]

\noindent Similarly, for add-noise $\rho$-QRs $\{F_t(X)\}_{t=1}^{n}$ as in~\eqref{eq:addnoisen}, we define
\begin{equation}
\label{eq:n-privacy-V}
\pi_{\rho}\left(V_1,\ldots,V_n\right)=P\big(g_{\text{MAP 
$\left(V_1,\ldots,V_n\right)$}}\left(Z_1,\ldots,Z_n\right)\neq X\big)
\end{equation} 

\noindent with
\[
g_{MAP\left(V_1,\ldots,V_n\right)}\left(i_1,\ldots,i_n\right)=\arg\hspace{0.1cm} 
\max_{x\in\cX}\hspace{1.5mm}
P_X\left(x\right)\prod\limits_{t=1}^{n}V_t\big(i_t|f(x)\big),\hspace{4mm} i_1,\ldots,i_n\in\cZ.
\]

\noindent Of particular interest will be the cases $W_t=W$ or $V_t=V,$ $t=1,\ldots,n,$ 
when we write~\eqref{eq:n-privacy-W} and~\eqref{eq:n-privacy-V} as
\[\pi_{\rho}\left(W^n\right)=P\big(g_{\text{MAP 
$\left(W^n\right)$}}\left(Z_1,\ldots,Z_n\right)\neq X\big)  \]

\noindent and 
\[\pi_{\rho}\left(V^n\right)=P\big(g_{\text{MAP 
$\left(V^n\right)$}}\left(Z_1,\ldots,Z_n\right)\neq X\big).\]
\end{definition}
\vspace{0.2cm}
We provide first in Section~\ref{sec:rhoprivn-converse} an upper bound for $\rho$-privacy $\pi_n(\rho)$ which 
is valid for each $0\leq\rho\leq 1$ and every $n\geq 1.$ Next, in Section~\ref{sec:rhoprivn-achievability}, 
considering the realms $0.5 < \rho \leq 1$ and $0\leq\rho\leq 0.5$ separately, we show corresponding explicit 
achievability schemes. However, unlike in Section~\ref{sec:rhopriv} for the case $n=1,$ the lower bound for 
$\pi_n(\rho)$ from the achievability schemes below, that use add-noise $\rho$-QRs, need not coincide with 
the upper bound in Theorem~\ref{thm:rhoprivn-converse} for any finite $n\geq 1.$ These upper and lower 
bounds for $\pi_n(\rho)$ are rendered into more convenient, albeit blunter forms in 
Section~\ref{sec:rhoprivn-bounds}.

\subsection{Converse}
\label{sec:rhoprivn-converse}
\noindent We provide next, as a converse result, an upper bound for $\pi_n(\rho),$ $n\geq 1$. 
For $0\leq\rho\leq 1,$ set

\begin{equation}
 \label{eq:gamma}
 \Gamma_n(\rho)=\min
\left\{1-\rho_c,\min\bigg\{1-\rho,P\left(\text{\normalfont{Bin}}
(n,\rho)\leq\left\lfloor\frac{n}{2}\right\rfloor\right)\bigg\}\right\}
\sum\limits_{i\in\cZ}P_X(x_i^*),\hspace{4mm}n\geq 1
\end{equation}

\noindent and note that $0\leq\Gamma_n(\rho)\leq 1.$

\begin{theorem}
\label{thm:rhoprivn-converse}
For each $0\leq\rho\leq 
1$ and for every $n\geq 1,$
\[ \pi_n(\rho) \leq 1-\sum\limits_{i\in\cZ}P_X(x_i^*)+\Gamma_n(\rho).\]
\end{theorem}

\noindent\textit{Remark}: For $0\leq\rho\leq 1$ and $n=1,$ since \[\Gamma_1(\rho)=
\left(1-\max\{\rho_c,\rho\}\right)
\sum\limits_{i\in\cZ}P_X\left(x_i^*\right),\] we have that the upper bound for $\pi_n(\rho)$ above reduces to 
that for $\pi(\rho)$ in the right-side of~\eqref{eq:thm-rhopriv-pf2}.
\vspace{0.2cm}\\
\noindent\textit{Proof}:\hspace{2mm}For $W_1,\ldots,W_n$ satisfying~\eqref{eq:recov2},
\begin{align}
P\big(g_{\text{MAP 
$\left(W_1,\ldots,W_n\right)$}}\left(Z_1,\ldots,Z_n\right)=X\big)&\geq 
P\big(g_{\text{MAP $\left(W_1\right)$}}\left(Z_1\right)=X\big)\nonumber\\
&\geq\max\left\{P_X\left(x^*\right),\rho\sum\limits_{i\in\cZ}
P_X\left(x_i^*\right)\right\}\label{eq:n-privacyub1}
\end{align}

\noindent by~\eqref{eq:thm-rhopriv-pf1}. Also,
\begin{equation}
\label{eq:n-privacyub2}
P\big(g_{\text{MAP $\left(W_1,\ldots,W_n\right)$}}\left(Z_1,\ldots,Z_n\right)= 
X\big)=\sum\limits_{\left(i_1,\ldots,i_n\right)\in\cZ^n}\max_{
x\in\cX}\hspace{1.5mm}P_X\left(x\right)\prod\limits_{t=1}^{n
}W_t\left(i_t|x\right).
\end{equation}

\noindent For each $i\in\cZ$ and for 
$l=\left\lfloor\frac{n}{2}\right\rfloor+1,\ldots,n,$ set
\begin{equation}
\label{eq:sign_terms}
\cA_l(i)=\left\{\left(i_1,\ldots,i_n\right)\in\cZ^n:\text{$i$ 
occurs $l$ times in $\left(i_1,\ldots, i_n\right)$}\right\}.
\end{equation}

\noindent Then, in~\eqref{eq:n-privacyub2},
\begin{align}
P\big(g_{\text{MAP $\left(W_1,\ldots,W_n\right)$}}\left(Z_1,\ldots,Z_n\right)= 
X\big)&\geq\sum\limits_{i\in\cZ}\sum\limits_{l=\left\lfloor\frac{n}{2}
\right\rfloor+1}^{n}\sum\limits_{\left(i_1,\ldots,i_n\right)\in\cA_l(i)}
\max_{x\in\cX}\hspace{1.5mm}
P_X\left(x\right)\prod\limits_{t=1}^{n}W_t\left(i_t|x\right)\nonumber\\ 
&\geq\sum\limits_{i\in\cZ}P_X\left(x_i^*\right)\sum\limits_{
l=\left\lfloor\frac{n}{2}\right\rfloor+1}^{n}\sum\limits_{\left(i_1,\ldots,
i_n\right)\in\cA_l(i)}\prod\limits_{t=1}^{n}
W_t\left(i_t|x_i^*\right)\nonumber\\
&=\sum\limits_{i\in\cZ}P_X\left(x_i^*\right)s_i(n)\label{eq:n-privacyub3}
\end{align}

\noindent where
\begin{equation}
\label{eq:n-privacyub4}
s_i(n)=\sum\limits_{l=\left\lfloor\frac{n}{2}\right\rfloor+1}^{n}s_i^l(n)
\end{equation}

\noindent with
\begin{equation}
\label{eq:n-privacyub5}
s_i^l(n)=\sum\limits_{\left(i_1,\ldots,i_n\right)\in\cA_l(i)}
\prod\limits_{t=1}^{n}W_t\left(i_t|x_i^*\right),\hspace{4mm}i\in\cZ.
\end{equation}

\noindent To understand the functional dependence of $s_i^l(n)$ on $\big(W_1\left(i|x_i^*\right),
\ldots,W_n\left(i|x_i^*\right)\big)$, consider as an instance all $\left(i_1,\ldots,i_n\right)\in\cA_l(i)$ 
with $i_1=\ldots=i_l=i$ and $i_t\neq i$, $t=l+1,\ldots,n$. The corresponding sum for such 
$\left(i_1,\ldots,i_n\right)\in\cA_l(i)$ in~\eqref{eq:n-privacyub5} equals
\[
\left(\prod\limits_{t=1}^{l}W_t\left(i|x_i^*\right)\right) 
\sum\limits_{\substack{\left(i_{l+1},\ldots,i_n\right)\in\cZ^{n-l} \\ i_t\neq i,\ t=l+1,\ldots,n}}\hspace{0.2cm} 
\prod\limits_{t=l+1}^{n}W_t\left(i_t|x_i^*\right)=\left(\prod\limits_{t=1}^{l}W_t\left(i|x_i^*\right)\right)
\prod\limits_{t=l+1}^{n}\left(1-W_t\left(i|x_i^*\right)\right).
\]

\noindent In this manner, we observe that $s_i^l(n)$ reduces to a sum of ${n \choose l}$ terms 
(corresponding to the locations of $l$ $i$s), each of which is a product of 
$W_t\left(i|x_i^*\right)$-terms for $l$ locations of $t$ in $\{1,\ldots,n\}$ corresponding 
to occurrences of $i,$ and $\big(1-W_t\left(i|x_i^*\right)\big)$-terms 
in the remaining $\left(n-l\right)$ locations. Thus, $s^l_i(n)$ is a function of 
$\big(W_1\left(i|x_i^*\right),\ldots,W_n\left(i|x_i^*\right)\big).$

\par We seek a suitable lower bound for $s_i(n)$ in terms of $\rho$ and $n$, to 
which end we make the
\vspace{0.1cm}\\
\textit{Claim}:\hspace{2mm}For $i\in\cZ,$ $s_i(n)$ is a nondecreasing 
function of each $W_t\left(i|x_i^*\right),$ $t=1,\ldots,n.$ 

\par By~\eqref{eq:n-privacyub4}, the claim and the observation 
following~\eqref{eq:n-privacyub5}, $s_i(n)$ is bounded below in an identical manner for 
$i=0,1,\ldots,k-1,$ upon replacing each $W_1\left(i|x_i^*\right),\ldots,
W_n\left(i|x_i^*\right)$ by $\rho,$ in accordance with~\eqref{eq:recov2}. By said observation, 
we have from~\eqref{eq:n-privacyub4} for $i=0,1,\ldots,k-1$ that 
\begin{align}
s_i(n)&\geq\sum\limits_{l=\left\lfloor\frac{n}{2}\right\rfloor+1}^{n}{n \choose 
l}\rho^l\left(1-\rho\right)^{n-l}\nonumber\\
&=P\left(\text{\normalfont{Bin}}(n,\rho)\geq\left\lfloor\frac{n}{2}
\right\rfloor+1\right)\label{eq:n-privacyub5a}.
\end{align}

\noindent Then from~\eqref{eq:n-privacyub3},
\begin{equation}\label{eq:n-privacyub6}
P\big(g_{\text{MAP 
$\left(W_1,\ldots,W_n\right)$}}\left(Z_1,\ldots,Z_n\right)=X\big)
\geq\left(\sum\limits_{i\in\cZ}P_X\left(x_i^*\right)\right)P\left(\text{\normalfont{Bin}}
(n,\rho)\geq\left\lfloor\frac{n}{2}\right\rfloor+1\right).
\end{equation}

\noindent Combining~\eqref{eq:n-privacyub1} and~\eqref{eq:n-privacyub6}, we get 
\begin{align*}
P\big(g_{\text{MAP 
$\left(W_1,\ldots,W_n\right)$}}\left(Z_1,\ldots,Z_n\right)=X\big)&\geq\max
\left\{P_X\left(x^*\right),\bigg(\sum\limits_{i\in\cZ}
P_X\left(x_i^*\right)\bigg)\max\bigg\{\rho,P\left(\text{
\normalfont{Bin}}(n,\rho)\geq\left\lfloor\frac{n}{2}\right\rfloor+1\right)\bigg\}
\right\} \\
&=\max\left\{\rho_c,\max\bigg\{\rho,P
\left(\text{\normalfont{Bin}}(n,\rho)\geq\left\lfloor\frac{n}{2}\right\rfloor+1\right)\bigg\}
\right\}\sum\limits_{i\in\cZ}P_X\left(x_i^*\right)\\
&=\sum\limits_{i\in\cZ}P_X\left(x_i^*\right)-\Gamma_n(\rho)
\end{align*}
\noindent from which the assertion of the theorem follows since $W_1,\ldots,W_n$ were 
arbitrary subject to~\eqref{eq:recov2}.

It remains to establish the claim, and it suffices to do so with $i=0,$ 
$t=1,$ i.e., we show that $s_0(n)$ is nondecreasing in 
$W_1\left(0|x_0^*\right)$. From the observation following~\eqref{eq:n-privacyub5}, 
$s_0^l(n)$ is a sum of ${n \choose l}$ terms, 
each of which is a product of $W_t\left(0|x_0^*\right)$-terms for $l$ locations 
of $t$ in $\{1,\ldots,n\}$ where $0s$ occur and 
$\big(1-W_t\left(0|x_0^*\right)\big)$-terms 
for the remaining $(n-l)$ locations. Thus, each of these ${n \choose l}$ terms 
will have either $W_1\left(0|x_0^*\right)$ or $1-W_1\left(0|x_0^*\right)$ in it 
(depending on whether or not $i_1=0$). The latter possibility yields a term 
with $-W_1\left(0|x_0^*\right)$ which is seen to be canceled by a suitable term 
from $s_0^{l+1}(n).$ Also, 
$s_0^n(n)=W_1\left(0|x_0^*\right)\prod\limits_{t=2}^n W_t\left(0|x_0^*\right).$ Thus, 
$s_0(n)$ consists of terms with $+W_1\left(0|x_0^*\right)$ or with no $W_1\left(0|x_0^*\right),$ 
and thereby is linear and nondecreasing in $W_1\left(0|x_0^*\right).$ This proves the claim.\qeed

\subsection{Achievability}
\label{sec:rhoprivn-achievability}
Throughout our achievability proofs, for the sake of convenience and without loss of 
essential generality, we assume that

\begin{equation}
\label{eq:prob-assumption}
P_X(x_i^*)\geq P_X(x_{i+1}^*),\hspace{4mm}i=0,1,\ldots,k-2. 
\end{equation}

\subsubsection{Realm $0.5<\rho\leq 1$}\hspace{1mm} 
\vspace{0.2cm}\\ Our achievability scheme uses 
the following stochastic matrix $V_1:\cZ\rightarrow\cZ$, not depending on $P_X$, given by
\vspace{0.2cm}
\begin{equation}
\label{eq:opt-scheme1}
 V_1(i|j)=\begin{cases}
           \rho,&i=j\\
           1-\rho,&\text{$j$ even and $i=j+1$ mod $k$ or $j$ odd and $i=j-1$}\\
           0,&\text{otherwise,}
          \end{cases}
\end{equation}

\noindent for $i,j \in \cZ.$ Thus, for $k$ even, the $k\times k$-matrix $V_1$ is block-diagonal 
with exactly $k/2$ blocks of $2\times 2$-matrices
\[
\begin{bmatrix}
\rho & 1-\rho\\
1-\rho & \rho
\end{bmatrix}
.\]

\noindent For $k$ odd, the upper-left $(k-1)\times (k-1)$-submatrix of $V_1$ is 
similarly structured with $(k-1)/2$ such blocks, and with the $k$th row being $V_1(0|k-1)=1-\rho$ 
and $V_1(k-1|k-1)=\rho.$ See Fig.~\ref{fig:matrix_V1}. Corresponding to $V_1:\cZ\rightarrow\cZ$ 
in~\eqref{eq:opt-scheme1}, consider the conditionally i.i.d. $\rho$-QRs $\{Z_t=F_t(X)\}_{t=1}^{n}$ 
given by~\eqref{eq:addnoisen} as
\begin{equation}
\label{eq:iid-QR}
P\big(F_1\left(X\right)=i_1,\ldots,F_n\left(X\right)=i_n\big|X=x\big)=\prod_{
t=1}^n V_1\left(i_t|f(x)\right).
\end{equation}

\begin{figure}
  \begin{subfigure}{0.5\textwidth}
  $$ 
\left[\begin{array}{cc:cc:c:cc}
\rho & 1-\rho &0 &0& \cdots&&\\
1 - \rho & \rho &0 &0& \cdots&&\\
\hdashline
0 & 0 &\rho & 1-\rho & \cdots&&\\
0 & 0 & 1 - \rho & \rho & \cdots&&\\
\hdashline
&&&&\ddots&&\\
\hdashline
&&\cdots&&&\rho&1-\rho\\
&&\cdots&&&1-\rho&\rho\\
\end{array}\right]
$$
    \caption{$k$ even}
  \end{subfigure}
  \hfill
  \begin{subfigure}{0.5\textwidth}
  $$ 
\left[\begin{array}{cc:cc:ccc:c}
\rho & 1-\rho &0 &0& \cdots&&&\\
1 - \rho & \rho &0 &0& \cdots&&&\\
\hdashline
0 & 0 &\rho & 1-\rho & \cdots&&&\\
0 & 0 & 1 - \rho & \rho & \cdots&&&\\
\hdashline
&&&&\ddots&&&\\
&&\cdots&&&\rho&1-\rho&0\\
0&0&\cdots&&&1-\rho&\rho&0\\
\hdashline
1-\rho&0&\cdots&&&&&\rho\\
\end{array}\right]
$$
\caption{$k$ odd}
\end{subfigure}
\caption{Add-noise $\rho$-QR $V_1.$}
\label{fig:matrix_V1}
\vspace{-0.5cm}
\end{figure}
\noindent For $0\leq\rho\leq 1,$ set 
\begin{equation}
\label{eq:lambda}
\Lambda_n(\rho)= P\left(\text{\normalfont{Bin}}(n,\rho)\leq\left\lfloor\frac{n}{2}\right\rfloor\right)
\left(\sum\limits_{i\in\cZ:\text{ $i$ \normalfont{odd}}}P_X\left(x_i^*\right)\right),\hspace{4mm}n\geq 1
\end{equation}

\noindent and note that $0\leq\Lambda_n(\rho)\leq 1.$

\begin{theorem}
\label{thm:rhoprivn-achiev1}
Let $0.5<\rho\leq 1.$ For every $n\geq 1,$ the add-noise $\rho$-QRs $\{Z_t=F_t(X)\}_{t=1}^{n}$ in~\eqref{eq:iid-QR} 
with $V_1:\cZ\rightarrow\cZ$ in~\eqref{eq:opt-scheme1} yield privacy
\begin{equation}
\label{eq:n-privacylb}
 \pi_\rho\left(V_1^n\right)\geq 1 - \sum\limits_{i\in\cZ}P_X\left(x_i^*\right) + \Lambda_n(\rho). 
\end{equation}

\end{theorem}

\noindent\textit{Remarks}:
\begin{enumerate}[(i)]
 \item The choice of $V_1:\cZ\rightarrow\cZ$ takes its cue from the proof of 
 Theorem~\ref{thm:rhoprivn-converse}. The first lower bound in~\eqref{eq:n-privacyub3} 
 results upon discarding those $(i_1,\ldots,i_n)$ in $\cZ^n$ in which the most frequent 
 symbol from $\cZ$ occurs no more than $\left\lfloor\frac{n}{2}\right\rfloor$ times. The specific 
choice of $V_1$ in~\eqref{eq:opt-scheme1} ensures that the number of such occurrences is at least 
$\left\lfloor\frac{n}{2}\right\rfloor+1$. 
\item Observe that when $P_X$ is the uniform pmf on $\cX$,
for $n = 1$, $\pi_{\rho}\left(V_1\right)=1-k\rho/r=\pi(\rho),$ 
the latter by~\eqref{eq:rhopriv}. On the other hand, $\pi_{\rho}\left(V_1\right)$ can be 
strictly smaller than $\pi(\rho);$ for instance for $\cX=\cZ=\{0,1,2\},$ $P_X=(0.5,0.3,0.2),$ 
$f(x)=x,$ and $\rho=0.6,$ it is straightforward to show that $\pi(\rho)=0.4$ whereas 
$\pi_{\rho}\left(V_1\right)=0.38.$
\end{enumerate}
\vspace{0.2cm}
\noindent\textit{Proof}:\hspace{2mm}We have
\begin{equation}
\label{eq:n-privacylb1}
 P\big(g_{\text{MAP$\left(V_1^n\right)$}}\left(Z_1,\ldots,Z_n\right)= 
X\big)=\sum\limits_{\left(i_1,\ldots,i_n\right)\in\cZ^n}\max_{
x\in\cX}\hspace{1.5mm}P_X(x)\prod\limits_{t=1}^{n}V_1\big(i_t|f(x)\big).
\end{equation}

\noindent When $\rho=1,$ $V_1:\cZ\rightarrow\cZ$ in~\eqref{eq:opt-scheme1} has $1$s 
along its diagonal and $0$s elsewhere. Hence, the right-side of~\eqref{eq:n-privacylb1} 
equals $\sum\limits_{i\in\cZ}P_X\left(x_i^*\right).$ Since $\Lambda_n(1)=0,$~\eqref{eq:n-privacylb} 
holds (with equality).
\par Hereafter we take $0.5<\rho<1.$ By the form of $V_1$ in~\eqref{eq:opt-scheme1}, for each 
$x\in\cX$ only those $\left(i_1,\ldots,i_n\right)\in\cZ^n$ yield nonzero contributions 
in~\eqref{eq:n-privacylb1} when consisting of $i_t=f(x);$ and $i_t=f(x)+1$ mod $k$ 
for $f(x)$ even or $i_t=f(x)-1$ for $f(x)$ odd. 
Accordingly, we distinguish between the cases when $k$ is even or it is odd.
\vspace{0.2cm}\\
(i) $k$ \textit{even}: For $i=0,2,\ldots,k-2,$ set
\begin{equation}
\label{eq:n-privacylb2}
\cB_n(i)=\big\{\left(i_1,\ldots,i_n\right)\in\cZ^n:\text{$i_t=i$ or $i_t=i+1$}\big\}.
\end{equation}

\noindent Then in~\eqref{eq:n-privacylb1}, 
\begin{equation}
\label{eq:n-privacylb3}
 P\big(g_{\text{MAP$\left(V_1^n\right)$}}\left(Z_1,\ldots,Z_n\right)= 
X\big)=\sum\limits_{i=0,2,\ldots,k-2}\sum\limits_{\left(i_1,\ldots,
i_n\right)\in\cB_n(i)}\max_{x\in f^{-1}(i)\cup 
f^{-1}(i+1)}\hspace{1.5mm}P_X(x)\prod\limits_{t=1}^{n}V_1\big(i_t|f(x)\big)
\end{equation}

\noindent where for each $i=0,2,\ldots,k-2$ and for each $\left(i_1,\ldots,i_n\right)\in\cB_n(i),$ 
\begin{multline}
\label{eq:n-privacylb4}
\max_{x\in f^{-1}(i)\cup f^{-1}(i+1)}\hspace{1.5mm}P_X(x)\prod
\limits_{t=1}^{n}V_1\big(i_t|f(x)\big)=\\
\max\left\{P_X(x_i^*)\rho^{l_i\left(i_1,\ldots,i_n\right)}\left(1-\rho\right)^{n-l_i\left(i_1,\ldots,i_n\right)},
P_X(x_{i+1}^*)\left(1-\rho\right)^{l_i\left(i_1,\ldots,i_n\right)}\rho^{n-l_i\left(i_1,\ldots,i_n\right)}\right\}
\end{multline}

\noindent with $l_i\left(i_1,\ldots,i_n\right)$ being the number of $i$s in $(i_1,\ldots,i_n).$ 
The first term in $\{\cdot,\cdot\}$ above is no larger than the second if 
\[l_i\left(i_1,\ldots,i_n\right)\leq\tau_n(i,\rho)\defn
\left\lfloor\frac{1}{2}\left(n-\frac{\text{log}\frac{P\left(x_{i}^*\right)}{P\left(x_{i+1}^*\right)}}
{\text{log}\frac{\rho}{1-\rho}}\right)\right\rfloor. \]

\noindent Since $0.5<\rho<1,$ we observe by the assumption 
in~\eqref{eq:prob-assumption} that $\tau_n(i,\rho)\leq\big\lfloor\frac{n}{2}\big\rfloor;$ 
and $\tau_n(i,\rho)\leq\frac{n}{2}-1$ for even\footnote{When $P\left(x_{i}^*\right)=P\left(x_{i+1}^*\right),$
 we get $\tau_n(i,\rho)=n/2$ for even $n.$ In this case, replacing $\tau_n(i,\rho)=n/2$ 
by $\tau_n(i,\rho)=n/2-1$ does not alter subsequent calculations.} $n$.

\par Then for $i=0,2,\ldots,k-2$ and when $\tau_n(i,\rho)\geq 0$, by~\eqref{eq:n-privacylb4} we get 
in~\eqref{eq:n-privacylb3} that
\begin{flalign}
&\sum\limits_{\left(i_1,\ldots,i_n\right)\in\cB_n(i)}\max_{x\in 
f^{-1}(i)\cup f^{-1}(i+1)}\hspace{1.5mm}P_X(x)\prod\limits_{t=1}^{n}V_1\big(i_t|f(x)\big)&\nonumber\\
&=P_X(x_{i+1}^*)\sum\limits_{l=0}^{\tau_n(i,\rho)}{n \choose l}\left(1-\rho\right)^l\rho^{n-l} + 
P_X(x_{i}^*)\sum\limits_{l=\tau_n(i,\rho)+1}^{n}{n \choose l}\rho^{l}\left(1-\rho\right)^{n-l}&\nonumber\\
&=P_X(x_{i+1}^*)\sum\limits_{l=n-\tau_n(i,\rho)}^{n}{n \choose l}\rho^{l}\left(1-\rho\right)^{n-l} + 
P_X(x_{i}^*)\sum\limits_{l=\tau_n(i,\rho)+1}^{n}{n \choose l}\rho^{l}\left(1-\rho\right)^{n-l}&\nonumber\\
&\leq P_X(x_{i+1}^*)\sum\limits_{l=\left\lfloor\frac{n}{2}\right\rfloor+1}^{n}{n \choose l}\rho^{l}
\left(1-\rho\right)^{n-l} 
+ P_X(x_{i}^*)\sum\limits_{l=0}^{n}{n \choose  l}\rho^{l}\left(1-\rho\right)^{n-l}&\label{eq:n-privacylb5}
\end{flalign}

\noindent where the first term in the previous inequality readily follows from the observation above, since 

\[n-\tau_n(i,\rho)\geq
\begin{cases}
&n-\left\lfloor\frac{n}{2}\right\rfloor\geq\left\lfloor\frac{n}{2}\right\rfloor+1\text{ for odd $n$}\\
&\frac{n}{2}+1\text{ for even $n$.}
\end{cases}\]

\noindent Note that when $\tau_n(i,\rho) <0$, this upper bound in~\eqref{eq:n-privacylb5} remains valid. 
By~\eqref{eq:n-privacylb3} and~\eqref{eq:n-privacylb5},
\begin{flalign}
&P\big(g_{\text{MAP$\left(V_1^n\right)$}}\left(Z_1,\ldots,Z_n\right)= X\big)\nonumber&\\
\leq & \sum\limits_{i=0,2,\ldots,k-2}\bigg[P_X\left(x_{i+1}^*\right)
P\left(\text{\normalfont{Bin}}
(n,\rho)\geq\left\lfloor\frac{n}{2}\right\rfloor+1\right) + 
P_X\left(x_{i}^*\right)P\left(\text{\normalfont{Bin}}
(n,\rho)\geq\left\lfloor\frac{n}{2}\right\rfloor+1\right)\nonumber&\\
&+P_X\left(x_{i}^*\right)P\left(\text{\normalfont{Bin}}(n,\rho)\leq\left\lfloor\frac{n}{2}
\right\rfloor\right)\bigg]\nonumber&\\
=&\left(\sum\limits_{i\in\cZ}P_X\left(x_i^*\right)\right)P\left(\text{\normalfont{Bin}}
(n,\rho)\geq\left\lfloor\frac{n}{2}\right\rfloor+1\right)+\left(\sum\limits_{i=0,2,\ldots,k-2}
P_X\left(x_i^*\right)\right)
P\left(\text{\normalfont{Bin}}(n,\rho)\leq\left\lfloor\frac{n}{2}\right\rfloor\right)
\label{eq:n-privacylb6}&\\
=&\sum\limits_{i\in\cZ}P_X\left(x_i^*\right)-\Lambda_n(\rho). \label{eq:n-privacylb7}&
\end{flalign}

\noindent (ii) $k$ \textit{odd}: For $i=0,2,\ldots,k-3,$ set $\cB_n(i)$ as in~\eqref{eq:n-privacylb2}, and
\begin{equation*}
\cB_n(k-1)=\big\{\left(i_1,\ldots,i_n\right)\in\cZ^n:\text{$i_t=0$ or $i_t=k-1$}\big\}.
\end{equation*}

\noindent Then
\begin{flalign}
&P\big(g_{\text{MAP$\left(V_1^n\right)$}}\left(Z_1,\ldots,Z_n\right)= X\big)\nonumber&\\
\leq&\sum\limits_{i=0,2,\ldots,k-3}\sum\limits_{\left(i_1,\ldots,i_n\right)\in\cB_n(i)}
\max_{x\in f^{-1}(i)\cup 
f^{-1}(i+1)}\hspace{1.5mm}P_X(x)\prod\limits_{t=1}^{n}V_1\big(i_t|f(x)\big)\label{eq:n-privacylb8-}&\\
&+
\sum\limits_{\left(i_1,\ldots,
i_n\right)\in\cB_n(k-1)}\max_{x\in f^{-1}(k-1)}\hspace{1.5mm}P_X(x)\prod\limits_{t=1}^{n}
V_1\big(i_t|f(x)\big)\nonumber&\\
\leq&\left(\sum\limits_{i=0}^{k-2}P_X\left(x_i^*\right)\right)
P\left(\text{\normalfont{Bin}}(n,\rho)\geq\left\lfloor\frac{n}{2}\right\rfloor+1\right)+
\sum\limits_{i=0,2,\ldots,k-3}P_X\left(x_i^*\right)P\left(\text{\normalfont{Bin}}(n,\rho)
\leq\left\lfloor\frac{n}{2}\right\rfloor\right)\nonumber&\\
&+P_X\left(x_{k-1}^*\right)\bigg(P\left(\text{\normalfont{Bin}}(n,\rho)
\geq\left\lfloor\frac{n}{2}\right\rfloor+1\right)
+P\left(\text{\normalfont{Bin}}(n,\rho)\leq\left\lfloor\frac{n}{2}\right\rfloor\right)\bigg)\nonumber&\nonumber&\\
=&\left(\sum\limits_{i\in\cZ}P_X\left(x_i^*\right)\right)P\left(\text{\normalfont{Bin}}(n,\rho)
\geq\left\lfloor\frac{n}{2}\right\rfloor+1\right)
+\left(\sum\limits_{i=0,2,\ldots,k-1}P_X\left(x_i^*\right)\right)P\left(\text{\normalfont{Bin}}
(n,\rho)\leq\left\lfloor\frac{n}{2}\right\rfloor\right)\nonumber&\\
=&\sum\limits_{i\in\cZ}P_X\left(x_i^*\right)-\Lambda_n(\rho) \label{eq:n-privacylb8}&
\end{flalign}

\noindent where in the inequality above, the first two terms on the right-side obtain 
\textit{a la}~\eqref{eq:n-privacylb6}. 
When $\left(i_1,\ldots,i_n\right)=(0,\ldots,0)$ (the all-zero sequence), the maximum 
in~\eqref{eq:n-privacylb8-} is over $x$ in $f^{-1}(0)\cup f^{-1}(1) \cup f^{-1}(k-1)$. 
The preceding calculations are, in effect over $x$ in $f^{-1}(0),$ and are justified 
since $P_X\left(x_0^*\right)\rho^n\geq  P_X\left(x_{1}^*\right)\left(1-\rho\right)^n 
\geq P_X\left(x_{k-1}^*\right)\left(1-\rho\right)^n$ for $0.5\leq\rho\leq 1.$ 
\par The assertion of the theorem holds by~\eqref{eq:n-privacylb7} and~\eqref{eq:n-privacylb8}.\qeed
\vspace{0.2cm}

\subsubsection{Realm $0\leq\rho\leq 0.5$}\hspace{1mm}  
\vspace{0.2cm}\\
Our achievability scheme uses $\rho$-QRs as in~\eqref{eq:iid-QR} with $V_1$ replaced by 
$V_2:\cZ\rightarrow\cZ$, not depending on $P_X$, which is: 
\\for $0\leq\rho\leq 1/k,$
\begin{equation}
 \label{eq:opt-scheme2-1}
 V_2(i|j)=\frac{1}{k},\hspace{4mm}i,j\in\cZ
\end{equation}
\noindent and for $1/k< \rho\leq 0.5,$
\begin{equation}
\label{eq:opt-scheme2-2}
 V_2(i|j)=\begin{cases}
           \frac{1}{\left\lfloor\frac{1}{\rho}\right\rfloor},&j=0,\ldots,\left\lfloor\frac{k}{\left\lfloor
           \frac{1}{\rho}\right\rfloor}\right\rfloor\left\lfloor\frac{1}{\rho}\right\rfloor-1,\hspace{2mm} 
           i=\left\lfloor\frac{j}{\left\lfloor\frac{1}{\rho}\right\rfloor}\right\rfloor\left\lfloor\frac{1}
           {\rho}\right\rfloor,\ldots,\left(\left\lfloor\frac{j}{\left\lfloor\frac{1}{\rho}\right\rfloor}\right
           \rfloor+1\right)\left\lfloor\frac{1}{\rho}\right\rfloor-1\\
           \frac{1}{k \text{ mod}\left\lfloor\frac{1}{\rho}\right\rfloor},&j=\left\lfloor\frac{k}
           {\left\lfloor\frac{1}{\rho}\right\rfloor}\right\rfloor\left\lfloor\frac{1}{\rho}\right\rfloor,
           \ldots,k-1,\hspace{2mm}i=\left\lfloor\frac{j}{\left\lfloor\frac{1}{\rho}\right\rfloor}\right
           \rfloor\left\lfloor\frac{1}{\rho}\right\rfloor,\ldots,k-1\\
           0,&\text{otherwise.}
          \end{cases}
\end{equation}

In particular, for $1/k<\rho\leq 0.5,$ the $k\times k$-matrix $V_2$ consists of 
$\left\lfloor\frac{k}{\left\lfloor\frac{1}{\rho}\right\rfloor}
\right\rfloor$ diagonal blocks of $\left\lfloor\frac{1} {\rho}\right\rfloor\times\left\lfloor
\frac{1}{\rho}\right\rfloor$-matrices, each with identical elements equal to $1/\left\lfloor
\frac{1}{\rho}\right\rfloor$; 
and a single ``filler'' block of size $k\text{ mod}\left\lfloor\frac{1} 
{\rho}\right\rfloor\times k\text{ mod}\left\lfloor\frac{1}{\rho}\right\rfloor$ with 
identical elements equal to $1/\left(k\text{ mod}\left\lfloor\frac{1}{\rho}\right\rfloor\right).$ 
The latter is vacuous if $\left\lfloor\frac{k}{\left\lfloor\frac{1}{\rho}\right\rfloor}
\right\rfloor\left\lfloor\frac{1}{\rho}\right\rfloor=k$, i.e., 
$k\text{ mod}\left\lfloor\frac{1}{\rho}\right\rfloor=0$. See Fig.~\ref{fig:matrxix_V2}.

\begin{figure}[h]
\centering
$$
\left[
\begin{array}{ccc:ccc:cc}
1/3&1/3&1/3&0&0&0&0&0\\
1/3&1/3&1/3&0&0&0&0&0\\
1/3&1/3&1/3&0&0&0&0&0\\
\hdashline
0&0&0&1/3&1/3&1/3&0&0\\
0&0&0&1/3&1/3&1/3&0&0\\
0&0&0&1/3&1/3&1/3&0&0\\
\hdashline
0&0&0&0&0&0&1/2&1/2\\
0&0&0&0&0&0&1/2&1/2\\
\end{array}
\right]
$$
\caption{Add-noise $\rho$-QR $V_2$ for $\rho=1/3$ and $k=8$.}
\label{fig:matrxix_V2}
\end{figure}

\begin{theorem}
\label{thm:rhoprivn-achiev2}
 Let $0\leq\rho\leq 0.5.$ For every $n\geq 1,$ the add-noise $\rho$-QRs $\left\{Z_t=F_t(X)\right\}_{t=1}^{n}$ 
 in~\eqref{eq:iid-QR} 
 with $V_1$ replaced by $V_2:\cZ\rightarrow\cZ$
 in~\eqref{eq:opt-scheme2-1},~\eqref{eq:opt-scheme2-2} yield privacy
\begin{equation}
\label{eq:thm_expression}
 \pi_\rho\left(V_2^n\right)=
 \begin{cases}
 1 - P_X\left(x^*\right), &0 \leq \rho \leq\frac{1}{k}\\
 1 - \sum\limits_{i=0}^{\left\lfloor\frac{k}{\left\lfloor\frac{1}{\rho}\right\rfloor}\right\rfloor}
       P_X\left(x_{i\left\lfloor\frac{1}{\rho}\right\rfloor}^*\right),&k\text{ \normalfont{mod}}
       \left\lfloor\frac{1}
{\rho}\right\rfloor\neq 0,\hspace{4mm}  \frac{1}{k} < \rho\leq 0.5\\
 1 - \sum\limits_{i=0}^{\left\lfloor\frac{k}{\left\lfloor\frac{1}{\rho}\right\rfloor}\right\rfloor-1}
       P_X\left(x_{i\left\lfloor\frac{1}{\rho}\right\rfloor}^*\right),&k\text{ \normalfont{mod}}
       \left\lfloor\frac{1}
{\rho}\right\rfloor=0,\hspace{4mm} \frac{1}{k} < \rho\leq 0.5.                                            
 \end{cases} 
\end{equation}

\end{theorem}
\vspace{0.2cm}

\noindent\textit{Remark}: For $0\leq\rho\leq 0.5,$ the privacy $\pi_\rho\left(V_2^n\right)$ 
above lacks dependence on $n.$ 
However, for $0\leq\rho\leq 1/k,$
\[\pi_\rho\left(V_2^n\right)=\pi_\rho\left(V_2\right)=1-P_X\left(x^*\right)=\pi(\rho)\]
where the last identity is by~\eqref{eq:rhopriv}. Thus, for $n=1,$ the add-noise $\rho$-QR with $V_2$ too 
achieves $\rho$-privacy, as did $V_o$ in Theorem~\ref{thm:rhopriv}. 

On the other hand, for $1/k<\rho\leq 0.5,$ $V_2$ can be strictly inferior to 
$V_o$ for $n=1;$ for instance, with $P_X$ being the uniform 
pmf on $\cX,$ by Theorem~\ref{thm:rhoprivn-achiev2} 
with $k\text{ mod}\left\lfloor\frac{1}{\rho}\right\rfloor\neq 0,$
\[\pi_\rho\left(V_2\right)=1-\left(\left\lfloor\frac{k}{\left\lfloor\frac{1}
{\rho}\right\rfloor}\right\rfloor+1\right)\frac{1}{r}
<1-\frac{k}{\left\lfloor\frac{1}{\rho}\right\rfloor}\frac{1}{r}\leq 
1-\frac{\rho k}{r}=\pi(\rho)=\pi(V_o) \]
where the last two identities are by Theorem~\ref{thm:rhopriv}.
\vspace{0.2cm}\\
\noindent\textit{Proof}:\hspace{2mm}We have
\begin{equation}
\label{eq:n-privacylb2_1}
 P\big(g_{\text{MAP$\left(V_2^n\right)$}}\left(Z_1,\ldots,Z_n\right)= 
X\big)=\sum\limits_{\left(i_1,\ldots,i_n\right)\in\cZ^n}\max_{
x\in\cX}\hspace{1.5mm}P_X(x)\prod\limits_{t=1}^{n}V_2\big(i_t|f(x)\big).
\end{equation}

\noindent When $0\leq \rho\leq 1/k,$ we get from~\eqref{eq:opt-scheme2-1} that 
\[P\big(g_{\text{MAP$\left(V_2^n\right)$}}\left(Z_1,\ldots,Z_n\right)= 
X\big)=P_X\left(x^*\right) \]
so that $\pi_\rho\left(V_2^n\right)=1-P_X\left(x^*\right).$ Considering next $1/k<\rho\leq 0.5,$ 
by the form of $V_2$ in~\eqref{eq:opt-scheme2-2}, for each 
\[x\in\cX\setminus\left\{f^{-1}\left(\left\lfloor\frac{k}{\left\lfloor\frac{1}
{\rho}\right\rfloor}\right\rfloor\left\lfloor\frac{1}{\rho}\right\rfloor\right)\cup\ldots\cup 
f^{-1}(k-1)\right\}\] 

\noindent only those $\left(i_1,\ldots,i_n\right)\in\cZ^n$ yield nonzero 
contributions in~\eqref{eq:n-privacylb2_1} when 
\[i_t\in\left\{\left\lfloor\frac{f(x)}
{\left\lfloor\frac{1}{\rho}\right\rfloor}\right\rfloor\left\lfloor\frac{1}{\rho}\right\rfloor,\ldots,
\left(\left\lfloor\frac{f(x)}{\left\lfloor\frac{1}{\rho}\right\rfloor}\right\rfloor+1\right)\left\lfloor
\frac{1}{\rho}\right\rfloor-1 \right\},\hspace{4mm} t=1,\ldots,n,\] 

\noindent and for each 
\[x\in \left\{f^{-1}\left(\left\lfloor\frac{k}{\left\lfloor\frac{1}{\rho}\right\rfloor}
\right\rfloor\left\lfloor\frac{1}{\rho}\right\rfloor
\right)\cup\ldots\cup f^{-1}(k-1)\right\} \]

\noindent only those $\left(i_1,\ldots,i_n\right)\in\cZ^n$ yield nonzero contributions 
in~\eqref{eq:n-privacylb2_1} when 
\[i_t\in\left\{\left\lfloor\frac{f(x)}{\left\lfloor\frac{1}{\rho}\right\rfloor}
\right\rfloor\left\lfloor\frac{1}{\rho}\right\rfloor,\ldots,k-1 \right\}, \hspace{4mm} t=1,\ldots,n.\]

\noindent For $i=0,\ldots,\left(\left\lfloor\frac{k}{\left\lfloor\frac{1}{\rho}\right\rfloor}\right\rfloor-1\right),$ set
\begin{equation*}
\cC_n(i)=\left\{\left(i_1,\ldots,i_n\right)\in\cZ^n:i_t\in\left\{i\left\lfloor\frac{1}{\rho}\right\rfloor,
\ldots,(i+1)\left\lfloor\frac{1}{\rho}\right\rfloor-1\right\}\right\}
\end{equation*}

\noindent and, when the filler block above exists,  
\[\cC_n\left(\left\lfloor\frac{k}{\left\lfloor\frac{1}{\rho}\right\rfloor}\right\rfloor\right)=
\left\{\left(i_1,\ldots,i_n\right)\in\cZ^n:i_t\in\left\{\left\lfloor\frac{k}{\left\lfloor\frac{1}{\rho}\right\rfloor}
\right\rfloor\left\lfloor\frac{1}{\rho}\right\rfloor,
\ldots,k-1\right\}\right\}.\]

\noindent Then in~\eqref{eq:n-privacylb2_1}, with the filler block existing
\begin{align}
 &P\left(g_{\text{MAP$\left(V_2^n\right)$}}\left(Z_1,\ldots,Z_n\right)=X\right)\nonumber\\
 =& \sum\limits_{i=0}^{\left\lfloor\frac{k}{\left\lfloor\frac{1}{\rho}\right\rfloor}\right\rfloor-1}
\sum\limits_{(i_1,\ldots,i_n)\in\cC_n(i)}\hspace{3mm}
 \max_{x\in f^{-1}\left(i\left\lfloor\frac{1}{\rho}\right\rfloor\right)\cup\ldots\cup 
 f^{-1}\left((i+1)\left\lfloor\frac{1}{\rho}\right\rfloor-1\right)}
\hspace{1.5mm}P_X(x)\prod\limits_{t=1}^{n}V_2\big(i_t|f(x)\big)\nonumber\\
 &+\sum\limits_{(i_1,\ldots,i_n)\in\cC_n\left(\left\lfloor\frac{k}{\left\lfloor\frac{1}{\rho}\right\rfloor}\right
 \rfloor\right)}\hspace{3mm}
 \max_{x\in f^{-1}\left(\left\lfloor\frac{k}{\left\lfloor\frac{1}{\rho}\right\rfloor}\right\rfloor\left\lfloor\frac{1}
{\rho}\right\rfloor\right)\cup\ldots\cup 
 f^{-1}\left(k-1\right) }\hspace{1.5mm}P_X(x)\prod\limits_{t=1}^{n}V_2\big(i_t|f(x)\big)\nonumber\\
 =& \sum\limits_{i=0}^{\left\lfloor\frac{k}{\left\lfloor\frac{1}{\rho}\right\rfloor}\right\rfloor-1}\sum
 \limits_{(i_1,\ldots,i_n)\in\cC_n(i)}\hspace{3mm}
 \max_{x\in f^{-1}\left(i\left\lfloor\frac{1}{\rho}\right\rfloor\right)\cup\ldots\cup f^{-1}\left((i+1)
 \left\lfloor\frac{1}{\rho}\right\rfloor-1\right)}
 \hspace{1.5mm}P_X(x)\left(\frac{1}{\left\lfloor\frac{1}{\rho}\right\rfloor}\right)^n\nonumber\\
 &+\sum\limits_{(i_1,\ldots,i_n)\in\cC_n\left(\left\lfloor\frac{k}{\left\lfloor\frac{1}{\rho}\right\rfloor}
 \right\rfloor\right)}\hspace{3mm}
 \max_{x\in f^{-1}\left(\left\lfloor\frac{k}{\left\lfloor\frac{1}{\rho}\right\rfloor}\right\rfloor\left
 \lfloor\frac{1}{\rho}\right\rfloor\right)\cup\ldots\cup 
 f^{-1}\left(k-1\right) }\hspace{1.5mm}P_X(x)\left(\frac{1}{k \text{ mod}
\left\lfloor\frac{1}{\rho}\right\rfloor}\right)^n\nonumber\\
 =& \sum\limits_{i=0}^{\left\lfloor\frac{k}{\left\lfloor\frac{1}{\rho}\right\rfloor}\right\rfloor-1}
 \left(\frac{1}{\left\lfloor\frac{1}{\rho}\right\rfloor}\right)^n
 \sum\limits_{(i_1,\ldots,i_n)\in\cC_n(i)}\hspace{3mm}
 \max_{x\in f^{-1}\left(i\left\lfloor\frac{1}{\rho}\right\rfloor\right)\cup\ldots\cup 
 f^{-1}\left((i+1)\left\lfloor\frac{1}{\rho}\right\rfloor-1\right)}
 \hspace{1.5mm}P_X(x)\nonumber\\
 &+\left(\frac{1}{k \text{ mod}\left\lfloor\frac{1}{\rho}\right\rfloor}\right)^n\sum
 \limits_{(i_1,\ldots,i_n)\in\cC_n\left(\left\lfloor\frac{k}{\left\lfloor
 \frac{1}{\rho}\right\rfloor}\right\rfloor\right)}\hspace{3mm}
 \max_{x\in f^{-1}\left(\left\lfloor\frac{k}{\left\lfloor\frac{1}{\rho}\right\rfloor}
 \right\rfloor\left\lfloor\frac{1}{\rho}\right\rfloor\right)\cup\ldots\cup 
 f^{-1}\left(k-1\right) }\hspace{1.5mm}P_X(x)\nonumber\\
 =& \sum\limits_{i=0}^{\left\lfloor\frac{k}{\left\lfloor\frac{1}{\rho}\right\rfloor}\right\rfloor-1}
 \left(\frac{1}{\left\lfloor\frac{1}{\rho}\right\rfloor}\right)^n
 \sum\limits_{(i_1,\ldots,i_n)\in\cC_n(i)} P_X\left(x^*_{i\left\lfloor\frac{1}{\rho}\right\rfloor}\right)
 +\left(\frac{1}{k \text{ mod}\left\lfloor\frac{1}{\rho}\right\rfloor}\right)^n
\sum\limits_{(i_1,\ldots,i_n)\in\cC_n\left(\left\lfloor\frac{k}{\left\lfloor\frac{1}{\rho}\right\rfloor}
\right\rfloor\right)} 
 P_X\left(x^*_{\left\lfloor\frac{k}{\left\lfloor\frac{1}{\rho}\right\rfloor}\right\rfloor\left\lfloor
 \frac{1}{\rho}\right\rfloor}\right)\label{eq:n-privacylb2_2}\\
 =& \sum\limits_{i=0}^{\left\lfloor\frac{k}{\left\lfloor\frac{1}{\rho}\right\rfloor}\right\rfloor}
 P_X\left(x_{i\left\lfloor\frac{1}{\rho}\right\rfloor}^*\right)\nonumber,
 \end{align}

\noindent where~\eqref{eq:n-privacylb2_2} uses~\eqref{eq:prob-assumption} and $\left\vert\cC_n(i)\right\vert=
\left(\left\lfloor\frac{1}{\rho}\right\rfloor\right)^n$ ,
$\left\vert\cC_n\left(\left\lfloor\frac{k}{\left\lfloor\frac{1}{\rho}
\right\rfloor}\right\rfloor\right)\right\vert=
\left(k \text{ mod}\left\lfloor\frac{1}{\rho}\right\rfloor\right)^n$. In the absence of 
the filler block, clearly 
\[P\left(g_{\text{MAP$\left(V_2^n\right)$}}\left(Z_1,\ldots,Z_n\right)=X\right)=
\sum\limits_{i=0}^{\left\lfloor\frac{k}{\left\lfloor\frac{1}{\rho}\right\rfloor}\right\rfloor-1}
P_X\left(x_{i\left\lfloor\frac{1}{\rho}\right\rfloor}^*\right).\]

\noindent The assertion of the theorem follows.\qeed

\subsection{Useful Bounds for $\pi_n(\rho)$}
\label{sec:rhoprivn-bounds}

Theorems~\ref{thm:rhoprivn-converse} and~\ref{thm:rhoprivn-achiev1} yield effective 
upper and lower bounds for $\pi_n(\rho).$ Upon rewriting these bounds with a 
slight weakening, useful information can be extracted concerning 
the limiting behaviour of $\pi_n(\rho)$ as $n\rightarrow\infty.$ Specifically by 
Theorem~\ref{thm:rhoprivn-converse}, for each $0\leq\rho\leq 1$ and for every $n\geq 1,$
\begin{equation}
 \label{eq:rhoprivn-ub1}
 \pi_n(\rho)\leq 1-\sum\limits_{i\in\cZ}P_X(x_i^*)+\Gamma_n(\rho)
\end{equation}

\noindent and by Theorem~\ref{thm:rhoprivn-achiev1}, for $0.5<\rho\leq 1$ and for every $n\geq 1,$
\begin{equation}
 \label{eq:rhoprivn-lb1}
 \pi_n(\rho)\geq\pi_{\rho}\left(V_1^n\right) 
\geq 1 - \sum\limits_{i\in\cZ}P_X\left(x_i^*\right) + \Lambda_n(\rho).
\end{equation}

\noindent Estimates of $P\left(\text{\normalfont{Bin}}(n,\rho)\leq\left\lfloor\frac{n}{2}\right\rfloor\right)$ 
appearing in $\Gamma_n(\rho)$ and $\Lambda_n(\rho)$ $\big($cf.~\eqref{eq:gamma} and~\eqref{eq:lambda}$\big)$ 
lead to useful bounds for $\pi_n(\rho)$ in~\eqref{eq:rhoprivn-ub1} and~\eqref{eq:rhoprivn-lb1}. 
Let Ber($\alpha$) denote a Bernoulli rv with the probability of ``1'' being $\alpha$, $0\leq\alpha\leq 1.$ 
Hereafter, all logarithms and exponentials are with respect to the base $2.$ 

\begin{lemma}
\label{lem:binom-bds}
\begin{enumerate}[(i)]
\item For each $0.5\leq\rho\leq 1$ and every $n\geq 1$,
\begin{multline*}
\frac{1}{n+1}
\exp\left[-nD\left(\text{\normalfont{Ber}$\left(\frac{1}{n}\left\lfloor\frac{n}{2}
\right\rfloor\right)$}\Big|\Big|\text{\normalfont{Ber}$\left(\rho\right)$}\right)\right]\leq 
P\left(\text{\normalfont{Bin}}(n,\rho)\leq\left\lfloor\frac{n}{2}\right\rfloor\right) 
\\\leq\left(\left\lfloor\frac{n}{2}\right\rfloor+1\right)\exp\left[-nD\left(\text{\normalfont{Ber}$
\left(\frac{1}{n}\left\lfloor\frac{n}{2}\right\rfloor\right)$}\Big|\Big|\text{\normalfont{Ber}$
\left(\rho\right)$}\right)\right].
\end{multline*}

\item For each $0\leq\rho\leq 0.5$ and for every $n\geq 1,$
\[P\left(\text{\normalfont{Bin}}(n,\rho)
\leq\left\lfloor\frac{n}{2}\right\rfloor\right)\geq 1-\rho.\]
\end{enumerate}
\end{lemma}

\noindent \textit{Proof}:\hspace{1mm}See Appendix~\ref{app:lemm_bds}. \qeed
\vspace{0.2cm}

\par Lemma~\ref{lem:binom-bds}(i) leads to the following useful bounds for $\pi_n(\rho).$ 

\begin{proposition}
\label{prop:rhoprivn-exp-bds}
For each $0.5<\rho\leq 1,$

\begin{enumerate}[(i)]
\item 

\begin{equation*}
\label{eq:rhoprivn-ub2}
\pi_n\left(\rho\right)\leq 1-\sum\limits_{i\in\cZ}P_X\left(x_i^*\right)+
\left(\left\lfloor\frac{n}{2}\right\rfloor+1\right)\exp\left[-nD\left(\text{\normalfont{Ber}$
\left(\frac{1}{n}\left\lfloor\frac{n}{2}\right\rfloor\right)$}\Big|\Big|\text{\normalfont{Ber}
$\left(\rho\right)$}\right)\right]\sum\limits_{i\in\cZ}P_X\left(x_i^*\right)
\end{equation*}

for all $n$ such that
\[\left(\left\lfloor\frac{n}{2}\right\rfloor+1\right)\exp\left[-nD\left(\text{\normalfont{Ber}$
\left(\frac{1}{n}\left\lfloor\frac{n}{2}\right\rfloor\right)$}\Big|\Big|\text{\normalfont{Ber}$
\left(\rho\right)$}\right)\right]\leq 1-\text{\normalfont{min}}\{\rho,\rho_c\}.\]
\item for every $n\geq 1,$
\begin{equation*}
\label{eq:rhoprivn-lb2}
\pi_n\left(\rho\right)\geq\pi_{\rho}\left(V_1^n\right) 
\geq 1-\sum\limits_{i\in\cZ}P_X\left(x_i^*\right)+\frac{1}{n+1}
\exp\left[-nD\left(\text{\normalfont{Ber}$\left(\frac{1}{n}\left\lfloor\frac{n}{2}
\right\rfloor\right)$}\Big|\Big|\text{\normalfont{Ber}$\left(\rho\right)$}\right)\right]
\left(\sum\limits_{i\in\cZ:\text{ $i$ \normalfont{odd}}}P_X\left(x_i^*\right)\right).
\end{equation*}
\end{enumerate}
\end{proposition}

\vspace{0.2cm}

\noindent\textit{Proof}: The assertions follow directly by applying the upper and lower 
bounds in Lemma~\ref{lem:binom-bds}(i) to the right-sides of~\eqref{eq:rhoprivn-ub1} 
and~\eqref{eq:rhoprivn-lb1}, respectively, and recalling~\eqref{eq:gamma} and~\eqref{eq:lambda}.\qeed

\subsection{Asymptotic Implications}

We close this section with useful asymptotic implications of 
Theorem~\ref{thm:rhoprivn-converse},\hspace{-0.5mm}~\ref{thm:rhoprivn-achiev1},\hspace{-0.2mm}~\ref{thm:rhoprivn-achiev2} 
and Proposition~\ref{prop:rhoprivn-exp-bds}. 
Considering first the (more interesting) realm $0.5<\rho\leq 1,$ the upper bounds for $\pi_n(\rho)$ in 
Theorem~\ref{thm:rhoprivn-converse} and Proposition~\ref{prop:rhoprivn-exp-bds}(i), as also the lower 
bounds in Theorem~\ref{thm:rhoprivn-achiev1} and Proposition~\ref{prop:rhoprivn-exp-bds}(ii), converge 
according to 
\begin{equation}
 \label{eq:rhoprivn-limit}
 \lim_{n} \pi_n(\rho) = 1-\sum\limits_{i\in\cZ} P_X\left(x_i^*\right)=\pi(1),\hspace{4mm} 0.5<\rho\leq 1
\end{equation}

\noindent (see Remark (ii) after Theorem~\ref{thm:rhopriv}), i.e., the error probability of a 
MAP estimator of $X$ on the basis of a knowledge of $f(X).$ Furthermore, both the sets 
of bounds converge at the same exponential rate in $n$ with the ($n$-dependent) exponent itself tending to 
$D\big(\text{\normalfont{Ber}$(0.5)$}||\text{\normalfont{Ber}$(\rho)$}\big)>0.$ 
Thus, in the realm $0.5<\rho\leq 1,$ the asymptotic privacy in~\eqref{eq:rhoprivn-limit} is 
that which is afforded when the querier forms an accurate MAP estimate of $f(X)$ w.p. $1$ 
from $\rho$-QRs $\left\{F_t(X)\right\}_{t=1}^n,$ followed by a MAP estimate of $X$ that is 
compatible with the estimated $f(X).$ 

\par In the realm $0\leq\rho\leq 0.5$, the upper bound for $\pi_n(\rho)$ in 
Theorem~\ref{thm:rhoprivn-converse}, by Lemma~\ref{lem:binom-bds}(ii), equals
\begin{equation}
\label{eq:lbasymptotic_implication}
1-\text{max}\{\rho_c,\rho\}\sum\limits_{i\in\cZ}P_X\left(x_i^*\right)
\end{equation}

\noindent for \textit{all} $n\geq 1,$ which is the $\rho$-privacy for $n=1$ in Theorem~\ref{thm:rhopriv}. 
As remarked after Theorem~\ref{thm:rhoprivn-achiev2}, this upper bound is 
unattainable, in general, by add-noise $\rho$-QRs $\left\{F_t(X)\right\}_{t=1}^n$ with $V_2:\cZ\rightarrow\cZ$
in~\eqref{eq:opt-scheme2-1},~\eqref{eq:opt-scheme2-2}. Hence, 
an interpretation as above in the complementary realm is lacking as is the answer to the 
putative tightness (or not) of the mentioned bound. However, since 
\[ \pi_n(\rho)\geq \pi_{\rho}\left(V_2^n\right)>1-\sum\limits_{i\in\cZ}P_X\left(x_i^*\right)=\pi(1),\]
where the strict inequality is evident from Theorem~\ref{thm:rhoprivn-achiev2} (by comparing the 
expressions in~\eqref{eq:thm_expression} with $1-\sum\limits_{i\in\cZ}P_X\left(x_i^*\right)$), 
we can conclude that no accurate estimate of $f(X)$ w.p. $1$ is possible from $\rho$-QRs $\{F_t(X)\}_{t=1}^n$ 
for any $n$, unlike for $0.5<\rho\leq 1$.



\section{Inadequacy of Conditionally i.i.d $W_o$ for Multiple Query Responses}
\label{sec:inadequacy-Wo}
Theorem~\ref{thm:rhopriv} establishes the optimality of the add-noise $\rho$-QR $W_o:\cX\rightarrow\cZ$, or 
equivalently $V_o:\cZ\rightarrow\cZ$, in achieving 
$\rho$-privacy $\pi(\rho),$ $0\leq\rho\leq 1,$ for $n=1.$ Upon choosing $W_t=W_o$ or $V_t=V_o$, $t=1,\ldots,n,$ $n\geq 2,$ 
in~\eqref{eq:n_recov} or~\eqref{eq:addnoisen}, respectively, how does the 
corresponding privacy $\pi_{\rho}\left(W_o^n\right)$ or $\pi_{\rho}\left(V_o^n\right)$ compare with 
the achievable privacy in Theorems~\ref{thm:rhoprivn-achiev1} and~\ref{thm:rhoprivn-achiev2}? 
In the regime of all suitably large $n,$ we show below that the former does not exceed the latter and, in fact, 
can be strictly smaller.

\par To this end, the concept of Chernoff information~\cite{Chernoff52} plays a material role. 
Given a stochastic matrix $V:\cZ\rightarrow\cZ$, define its \textit{Chernoff radius}, 
denoted $C(V)$, as the minimum of pairwise Chernoff information quantities:
\begin{align}
C(V)&=\min_{\substack{j\neq j'\\ j,j'\in\cZ}}C(j,j')\nonumber\\
\label{eq:chern-rad}
    &=\min_{\substack{j\neq j'\\ j,j'\in\cZ}}\left[-\min_{0\leq\lambda\leq 1}\log\left(\sum_{i\in\cZ}
    V(i|j)^{\lambda}V\left(i|j'\right)^{1-\lambda}\right)\right],
\end{align}
noting that $C (V) \geq 0$ with $C (V) > 0$ iff all the rows of $V$ are distinct.

Also useful will be the next two technical lemmas. Let $\tilde{f}(X)$ be a $\cZ$-valued rv with pmf
\begin{equation*}
P\left(\tilde{f}(X)=i\right)=\frac{P_X(x_i^*)}{\sum\limits_{l\in\cZ}P_X\left(x_l^*\right)},\hspace{4mm}i \in\cZ
\end{equation*}

\noindent with $x_i^*,$ $i\in\cZ,$ as in~\eqref{eq:prob}. Let $\tilde{Z}_t,$ $t=1,\ldots, n$, 
be conditionally mutually independent $\cZ$-valued rvs conditioned on $\tilde{f}(X),$ with
\begin{equation*}
 P_{\tilde{Z}_t|\tilde{f}(X)}=V,\hspace{4mm}t=1,\ldots,n.
\end{equation*}

\noindent We use the notation $A\doteq\exp(-nB)$ to mean $\displaystyle{\lim_{n}\hspace{1mm}-\frac{1}{n}\log A=B}$ 
(cf. e.g., \cite{KanayaHan95}).

\begin{lemma}
\label{lem:privacy-VW-equiv}
For $0\leq\rho\leq 1,$ consider add-noise $\rho$-QRs $\left\{F_t(X)\right\}_{t=1}^{\infty}$ with~\eqref{eq:n_recov} 
holding for every $n\geq 1,$ where 
$W_t=W,$ $t\geq 1,$ and $W:\cX\rightarrow\cZ$ has identical rows for all $x\in f^{-1}(i),$ $i\in\cZ,$ and 
has associated $V:\cZ\rightarrow\cZ$ in~\eqref{eq:lem-addnoise1}.
\begin{enumerate}[(i)]
 \item The corresponding privacy for every $n\geq 1$ is

 \begin{equation}
 \label{eq:lemma_VW_equivalence}
\pi_{\rho}\left(W^n\right) = \pi_{\rho}\left(V^n\right) = 1 - \left(\sum_{i\in\cZ}P_X\left(x_i^*\right)\right)
P\left(g_{\text{\normalfont{MAP}
$\left(V^n\right)$}}\left(\tilde{Z_1},\ldots,\tilde{Z_n}\right)=\tilde{f}(X)\right).
\end{equation}

\item Furthermore,

\begin{equation}
 \label{eq:lemm_chernoff}
 \pi_{\rho}\left(V^n\right) - \left(1-\sum_{i\in\cZ}P_X\left(x_i^*\right)\right)\doteq\exp\left[-nC(V)\right].
\end{equation}

\end{enumerate}
\end{lemma}
\vspace{0.2cm}
\noindent \textit{Proof}:
\begin{enumerate}[(i)]
 \item 
\begin{align}
 P\left(g_{\text{MAP$\left(W^n\right)$}}\left(Z_1,\ldots,Z_n\right)=X\right)
 &=\sum\limits_{\left(i_1,\ldots,i_n\right)\in\cZ^n}\max_{
x\in\cX}\hspace{1.5mm}P_X(x)\prod\limits_{t=1}^{n}W\left(i_t|x\right)\nonumber\\
&=\sum\limits_{\left(i_1,\ldots,i_n\right)\in\cZ^n}\max_{\substack{x\in\cup f^{-1}(j) \\ j\in\cZ}}
\hspace{1.5mm}P_X(x)\prod\limits_{t=1}^{n}W\left(i_t|x\right)\nonumber\\
&=\sum\limits_{\left(i_1,\ldots,i_n\right)\in\cZ^n}\max_{
j\in\cZ}\hspace{1.5mm}P_X\left(x_j^*\right)\prod\limits_{t=1}^{n}
W\left(i_t|x_j^*\right)\hspace{4mm}\nonumber\\
&=\sum\limits_{\left(i_1,\ldots,i_n\right)\in\cZ^n}\max_{
j\in\cZ}\hspace{1.5mm}P_X\left(x_j^*\right)\prod\limits_{t=1}^{n}V\left(i_t|f\left(x_j^*\right)\right),
\hspace{4mm}\text{by~\eqref{eq:lem-addnoise1}}\nonumber\\
&=\sum\limits_{\left(i_1,\ldots,i_n\right)\in\cZ^n}\max_{
j\in\cZ}\hspace{1.5mm}P_X(x_j^*)\prod\limits_{t=1}^{n}V\left(i_t|j\right)\label{eq:privacy-VW-equiv}\\
&=\left(\sum\limits_{i\in\cZ}P_X\left(x_i^*\right)\right)\sum\limits_{\left(i_1,\ldots,i_n\right)\in\cZ^n}
\max_{j\in\cZ}\hspace{1.5mm}\frac{P_X\big(x_j^*\big)}{\sum\limits_{i\in\cZ}P_X\left(x_i^*\right)}
\prod\limits_{t=1}^{n}V\left(i_t|j\right)\nonumber\\
&=\left(\sum\limits_{i\in\cZ}P_X\left(x_i^*\right)\right) P\left(g_{\text{MAP$\left(V^n\right)$}}
\left(\tilde{Z}_1,\ldots,\tilde{Z}_n\right)=\tilde{f}(X)\right)\label{eq:privacy-VW-equiv-proof}
\end{align}

where the third equality above is by the assumed form of $W.$ 
\par The first assertion in~\eqref{eq:lemma_VW_equivalence} follows from~\eqref{eq:privacy-VW-equiv} 
and the second from~\eqref{eq:privacy-VW-equiv-proof}.

\item By {\cite[Theorem $2$]{KanayaHan95}}, 
\[P\left(g_{\text{MAP$\left(V^n\right)$}}\left(\tilde{Z}_1,\ldots,\tilde{Z}_n\right)
\neq\tilde{f}(X)\right)\doteq \exp\left[-nC(V)\right], \]
which, applied to~\eqref{eq:lemma_VW_equivalence}, yields~\eqref{eq:lemm_chernoff}.
\end{enumerate} \qeed\\

\noindent\textit{Remark}: Observe that a direct application of {\cite[Theorem $2$]{KanayaHan95}} to
\[\pi_{\rho}\left(W^n\right)= P\left(g_{\text{MAP$\left(W^n\right)$}}\left(Z_1,\ldots,Z_n\right)\neq X\right)\]
is not useful as it yields
\[\pi_{\rho}\left(W^n\right)\doteq\exp\left[-nC(W)\right]\]
where the Chernoff radius of $W:\cX\rightarrow\cZ$ is $C(W)=0$ owing to the presence of identical rows when 
$k\leq r-1$.
\begin{lemma}
\label{lem:chernoff-compare}
For $V_o$ in~\eqref{eq:optchannel2} and $V_1$ in~\eqref{eq:opt-scheme1}, we have
\begin{equation}
\label{eq:lemm_chernoff_compare1}
C\left(V_1\right)=D\big(\text{\normalfont{Ber}}(0.5)||\text{\normalfont{Ber}}(\rho)\big)= 
-\log2\sqrt{\rho(1-\rho)},\hspace{4mm}0\leq\rho\leq 1
\end{equation}
 \noindent and for $0.5<\rho<1$
\begin{align}
 C\left(V_o\right) &= -\log2\sqrt{\max\{\rho_c,\rho\}\left(1-\max\{\rho_c,\rho\}\right)},
 \hspace{4mm}k=2\label{eq:lemm_chernoff_compare2}\\
 C\left(V_o\right) &> -\log2\sqrt{\max\{\rho_c,\rho\}\left(1-\max\{\rho_c,\rho\}\right)},
 \hspace{4mm}k\geq 3.\label{eq:lemm_chernoff_compare3}
\end{align}
\end{lemma}
\noindent \textit{Proof}:\hspace{1mm} First observe that for $0<\rho<1,$
\begin{align}
C\left(V_1\right)
&=\sup_{0< \lambda< 1}\log\frac{1}{\rho^{\lambda}\left(1-\rho\right)^{1-\lambda}+
\rho^{1-\lambda}\left(1-\rho\right)^{\lambda}}\nonumber\\
&=\log\frac{1}{\underset{0< \lambda< 1}{\text{inf}}\rho^{\lambda}\left(1-\rho\right)^{1-\lambda}+
\rho^{1-\lambda}\left(1-\rho\right)^{\lambda}}\nonumber\\
&=\log\frac{1}{2\sqrt{\rho(1-\rho)}}\label{eq:lem-cher-proof0}\\
&=D\big(\text{\normalfont{Ber}}(0.5)||\text{\normalfont{Ber}}(\rho)\big)\nonumber
\end{align}
where the infimum is attained as a minimum at $\lambda=0.5;$ and $C(V_1)=\infty$ for $\rho=0$ and $\rho=1.$ 
The last equality above is by simple calculation.
\par Turning to~\eqref{eq:lemm_chernoff_compare2}, for $k=2,$
\begin{align*}
C\left(V_o\right)
&=\sup_{0< \lambda< 1}\log\frac{1}{\left(\max\{\rho_c,\rho\}\right)^{\lambda}
\left(1-\max\{\rho_c,\rho\}\right)^{1-\lambda}+
\left(\max\{\rho_c,\rho\}\right)^{1-\lambda}\left(1-\max\{\rho_c,\rho\}\right)^{\lambda}}\\
&=\log\frac{1}{2\sqrt{\max\{\rho_c,\rho\}\left(1-\max\{\rho_c,\rho\}\right)}},
\end{align*}

\noindent in the manner of~\eqref{eq:lem-cher-proof0}.
\par To show~\eqref{eq:lemm_chernoff_compare3}, for $j\neq j'$ in $\cZ,$
\begin{align}
C\left(j,j'\right)&=\sup_{0< \lambda< 1}\log\frac{1}{\sum\limits_{i\in\cZ}
V_o(i|j)^{\lambda}V_o(i|j')^{1-\lambda}}\nonumber\\
&=\sup_{0< \lambda< 1}\left(1-\lambda\right)D_{\lambda}\big(V_o(.|j)||V_o(.|j')\big)
\label{eq:lem-cher-proof2}
\end{align} 

\noindent where $D_{\lambda}$ is the R{\'e}nyi divergence of order $\lambda$~\cite{Renyi61}.
For each $\lambda\in(0,1),$ since $D_{\lambda}$ satisfies the data processing theorem 
{\cite[Theorem $1$]{ErvenHarremoes14}}, we get
\begin{align}
D_{\lambda}\big(V_o(.|j)||V_o(.|j')\big) 
&\geq D_{\lambda}\left(\text{Ber}\left(V_o(j'|j)\right)||\text{Ber}\left(V_o(j'|j')\right)\right)\nonumber\\
&= D_{\lambda}\left(\text{Ber}
\left(\frac{P_X\big(x_{j'}^*\big)}{\sum\limits_{l\neq j}P_X\left(x_{l}^*\right)}
\big(1-\max\{\rho_c,\rho\}\big) \right)
\Big|\Big|\text{Ber}\big(\max\{\rho_c,\rho\}\big)\right)\label{eq:lem-cher-proof3}.
\end{align}

\noindent \textit{Claim}: For $k\geq 3$ and $0.5<\rho<1,$ the right-side of~\eqref{eq:lem-cher-proof3} 
is strictly larger than 
\begin{equation*}
D_{\lambda}\left(\text{Ber}
\left(1-\max\{\rho_c,\rho\} \right)
\big|\big| \text{Ber}\big(\max\{\rho_c,\rho\}\big)\right).
\end{equation*} 

\noindent Then applying the claim to~\eqref{eq:lem-cher-proof2}, for all $j\neq j'$ in $\cZ,$
\begin{align*}
C(j,j')&>\sup_{0< \lambda< 1}(1-\lambda)D_{\lambda}\left(\text{Ber}
\left(1-\max\{\rho_c,\rho\} \right)
\big|\big|\text{Ber}\big(\max\{\rho_c,\rho\}\big)\right)\\
 &\geq 0.5 D_{0.5}\left(\text{Ber}
\left(1-\max\{\rho_c,\rho\} \right)
\big|\big|\text{Ber}\big(\max\{\rho_c,\rho\}\big)\right)\\
 &=-\log2\sqrt{\max\{\rho_c,\rho\}(1-\max\{\rho_c,\rho\})}
\end{align*}

\noindent which yields~\eqref{eq:lemm_chernoff_compare3}. 
\par It remains to prove the claim. Note that for $k\geq 3,$ $P_X\big(x_{j'}^*\big)\Big/
\sum\limits_{l\neq j}P_X\left(x_{l}^*\right) <1$ and so 
\[\frac{P_X\big(x_{j'}^*\big)}{\sum\limits_{l\neq j}P_X\left(x_{l}^*\right)}\big(1-\max\{\rho_c,\rho\}\big) 
< 1- \max\{\rho_c,\rho\}<\max\{\rho_c,\rho\}\]

\noindent since $\max\{\rho_c,\rho\}>0.5.$ Then, it suffices to show that 
$D_{\lambda}\big(\text{Ber}(\alpha)||\text{Ber}(\beta)\big)$ is 
(strictly) decreasing in $\alpha$ for $0\leq\alpha< \beta.$ 
We have

\begin{equation}
\label{eq:chern-diff}
\frac{d}{d\alpha}D_{\lambda}\big(\text{Ber}(\alpha)||\text{Ber}(\beta)\big)=\frac{1}{\lambda-1}
\frac{\lambda\alpha^{\lambda-1}\beta^{1-\lambda}-\lambda\left(1-\alpha\right)^{\lambda-1}
\left(1-\beta\right)^{1-\lambda}}{\alpha^{\lambda}\beta^{1-\lambda}+\left(1-\alpha\right)^{\lambda}
\left(1-\beta\right)^{1-\lambda}}.
\end{equation}

\noindent Since $\lambda\in(0,1),$ the right-side of~\eqref{eq:chern-diff} is negative iff 

\begin{equation*}
\alpha^{\lambda-1}\beta^{1-\lambda} > \left(1-\alpha\right)^{\lambda-1}
\left(1-\beta\right)^{1-\lambda}, \ {\rm i.e.,} \
\left(\frac{1-\alpha}{\alpha}\right)^{1-\lambda} > \left(\frac{1-\beta}{\beta}\right)^{1-\lambda}
\end{equation*}

\noindent 
which holds since $\alpha<\beta$.
\qeed

\vspace{0.2cm}
\par Finally, we show that the privacy of add-noise $\rho$-QRs $\{F_t(X)\}_{t=1}^{\infty}$ 
under~\eqref{eq:addnoisen} for every $n\geq 1$ with $V_t=V_o,$ $t\geq 1,$ is no 
better than with $V_t=V_1$ or $V_2$ accordingly as 
$0.5<\rho\leq 1$ or $0\leq\rho\leq 0.5;$ and, in fact, the former 
can be strictly smaller than the latter. 

\begin{proposition}
\label{prop:Vo-chernoff}
For all $n$ suitably large (depending on case below):
\begin{enumerate}[(i)]
\item $0\leq \rho\leq 0.5$\normalfont{:}
\begin{equation}
\label{eq:prop-chernoff0}
 \pi_{\rho}\left(V_2^n\right)\geq \pi_{\rho}\left(V_o^n\right); 
\end{equation}

\item $0.5< \rho< 1$\normalfont{:}\\
$k=2$\hspace{4.5mm}$-$  
\begin{align}
\pi_{\rho}\left(V_1^n\right)>\pi_{\rho}\left(V_o^n\right),
&\hspace{4mm}\rho<\rho_c\label{eq:prop-chernoff1}\\
\pi_{\rho}\left(V_1^n\right)=\pi_{\rho}\left(V_o^n\right),
&\hspace{4mm}\rho\geq\rho_c;\label{eq:prop-chernoff2}
\end{align}

$k\geq 3$\hspace{4.5mm}$-$  
\begin{equation}
\label{eq:prop-chernoff3}
\pi_{\rho}\left(V_1^n\right)>\pi_{\rho}\left(V_o^n\right).
\end{equation}

\end{enumerate}
\end{proposition}
\vspace{0.2cm}

\noindent\textit{Proof}:\hspace{1mm}
\begin{enumerate}[(i)]
\item See Appendix~\ref{app:prop_chernoff}.
\item For $0\leq \rho< 1,$ we have by Lemma~\ref{lem:privacy-VW-equiv}(ii),
\begin{equation}
\label{eq:prop_chernoff_compare1}
\pi_{\rho}\left(V_o^n\right)-\left(1-\sum_{i\in\cZ}P_X\left(x_i^*\right)\right) 
\doteq \exp\left[-nC\left(V_o\right)\right],
\end{equation}

and by Theorem~\ref{thm:rhoprivn-achiev1} for $0.5<\rho \leq 1,$ 
\begin{align}
\pi_{\rho}\left(V_1^n\right)-\left(1-\sum_{i\in\cZ}P_X\left(x_i^*\right)\right) &\geq \Lambda_n(\rho),
\hspace{4mm}n\geq 1\nonumber\\
 &\doteq\exp\big[-nD\big(\text{\normalfont{Ber}}(0.5)||\text{\normalfont{Ber}}(\rho)\big)\big],\hspace{2mm}
 \text{by~\eqref{eq:lambda} and Lemma~\ref{lem:binom-bds}(i)} \nonumber\\
 &=\exp\big[-nC(V_1)\big],\hspace{2mm}\text{by~\eqref{eq:lemm_chernoff_compare1}}\label{eq:prop_chernoff_compare2}.
\end{align}

\noindent For $k=2$ and $0.5<\rho<\rho_c,$ by~\eqref{eq:lemm_chernoff_compare1} 
and~\eqref{eq:lemm_chernoff_compare2},
\begin{equation*}
C(V_1)=-\log{2\sqrt{\rho(1-\rho)}}<-\log{2\sqrt{\rho_c(1-\rho_c)}}=C(V_o)
\end{equation*}
so that~\eqref{eq:prop-chernoff1} holds by~\eqref{eq:prop_chernoff_compare1} 
and~\eqref{eq:prop_chernoff_compare2}. 
For $k=2$ and $\rho\geq\rho_c,$ observe in~\eqref{eq:opt-scheme1} and~\eqref{eq:optchannel2}
that $V_1=V_o$
whereby~\eqref{eq:prop-chernoff2} holds.
For $k\geq 3$ and $0.5<\rho<1,$ by~\eqref{eq:lemm_chernoff_compare1} and~\eqref{eq:lemm_chernoff_compare3},

\[C\left(V_1\right)=-\log{2\sqrt{\rho(1-\rho)}}\leq 
-\log2\sqrt{\max\{\rho_c,\rho\}\left(1-\max\{\rho_c,\rho\}\right)}<C(V_o)\]

and so~\eqref{eq:prop-chernoff3} holds by~\eqref{eq:prop_chernoff_compare1} 
and~\eqref{eq:prop_chernoff_compare2}.
\end{enumerate}\qeed



\section{Discussion}
\label{sec:discussion}

The choice of $W_o:\cX\rightarrow\cZ$ or $V_o:\cZ\rightarrow\cZ$ 
in~\eqref{eq:optchannel1},~\eqref{eq:optchannel2}, depending on 
$P_X$ through $P_X\left(x_i^*\right),$ $i \in \cZ$,
yields maximal privacy for a single $\rho$-QR for all $0\leq\rho\leq 1$. However, for the case 
of multiple conditionally independent $\rho$-QRs, our achievability schemes in Section~\ref{sec:rhoprivn},
that are ``universal'' in the sense of not depending on $P_X$, perform variously according to 
the value of $\rho$. In particular, for $0.5<\rho\leq 1$, conditionally i.i.d. 
add-noise $\rho$-QRs $\left\{F_t(X)\right\}_{t=1}^{\infty}$ with $V_1:\cZ\rightarrow\cZ$ 
in~\eqref{eq:opt-scheme1} are asymptotically optimal with privacy $\pi_{\rho}\left(V_1^n\right)$ 
converging to the limit of the upper bounds for $\rho$-privacy $\pi_n(\rho)$, $n\geq 1$, 
in Theorem~\ref{thm:rhoprivn-converse}. However, when $0\leq\rho\leq 0.5$, our add-noise 
$\rho$-QRs with $V_2:\cZ\rightarrow\cZ$ in~\eqref{eq:opt-scheme2-1},~\eqref{eq:opt-scheme2-2} 
yield privacy $\pi_{\rho}\left(V_2^n\right)$ not depending on $n$, which, in general, 
does not meet the corresponding upper bound in~\eqref{eq:lbasymptotic_implication}. 
Thus, it remains open whether conditionally independent $\rho$-QRs $\left\{F_t(X)\right\}_{t=1}^{\infty}$, 
that depend on $P_X$ or are not necessarily of the add-noise variety, 
can outperform $\pi_{\rho}\left(V_1^n\right)$ or $\pi_{\rho}\left(V_2^n\right)$. 
Indeed, the goodness of our upper bound for $\pi_n(\rho)$ in~\eqref{eq:lbasymptotic_implication}, 
$0\leq\rho\leq 0.5$ (that does not depend on $n$), is unresolved. These observations are analogous -- 
in our setting -- to the ``composition'' results for differential privacy (cf. e.g.,~\cite{Kairouz17}).

\par We conclude with a simple observation in explication of our approach mentioned in 
Section~\ref{sec:intro}. Suppose that the querier's family of priors $\mathcal{P}$ consists of a 
specified set of pmfs $P$ on $\cX$ with $P_X(x)>0$, $x\in\cX$. For a single $\rho$-QR, the 
$\rho$-privacy $\pi(\rho)=\pi(\rho;P)$ for any $P$ in $\mathcal{P}$ is attained by $W_o=W_o(P)$ or 
$V_o=V_o(P)$ as remarked after Theorem~\ref{thm:rhopriv}. 
With
\[P_*=P_*(\rho) = \arg\hspace{0.1cm}\min_{P\in\mathcal{P}}\hspace{1.5mm}\pi(\rho;P),\hspace{4mm}0\leq\rho\leq 1\]

\noindent a $\rho$-QR $W_o(P_*)$ or $V_o(P_*)$ will yield privacy $\pi(\rho;P_*)$ in~\eqref{eq:rhopriv} 
that serves as a guaranteed lower bound for $\rho$-privacy computed according to \textit{any} prior pmf 
$P$ in $\mathcal{P}$. In the same vein, for $n\geq 1$ conditionally independent query responses, the minima with respect 
to $P$ in $\mathcal{P}$ of the lower bound for $\pi_{\rho}\left(V_1^n\right)$ in~\eqref{eq:n-privacylb} or of 
$\pi_{\rho}\left(V_2^n\right)$ in Theorem~\ref{thm:rhoprivn-achiev2}, respectively, serve as privacy guarantees 
in the realms $0.5<\rho\leq 1$ or $0\leq \rho\leq 0.5$, computed for any $P$ in $\mathcal{P}$.

\ifx
\par We conclude with a discussion of the $(\epsilon,\delta)$-differential privacy afforded 
by the $\rho$-QRs in Sections~\ref{sec:rhopriv} and~\ref{sec:rhoprivn} above. In our context, 
a $\rho$-QR $F(X)$ with $W:\cX\rightarrow\cZ$ is $(\epsilon,\delta)$-differential private, $\epsilon\geq 0$, 
$0\leq\delta\leq 1$, if for all $x\neq x'$ in $\cX$, it holds that~\cite{Dwork06},
\[W(S|x)\leq e^{\epsilon}W(S|x')+\delta,\hspace{4mm}S \subseteq
\cX.\]
Considering first the $\rho$-QR $W_o:\cX\rightarrow\cZ$ in~\eqref{eq:optchannel1}, it suffices 
to consider 
$x\neq x'$ in $\cX$ with  $f(x)\neq f(x')$. Given $0\leq\delta\leq 1$, $(\epsilon,\delta)$-differential 
privacy is attainable by $W_o$ for all $\epsilon\geq\epsilon^*=\epsilon^*(\delta,\rho,P_X)$ with
\begin{equation}
\label{eq:eps_diff_privacy}
\epsilon^*(\delta,\rho,P_X)=\max_{\substack{x,x':x\neq x',f(x)\neq f(x')\\ S\subseteq \cX}} 
\left(\ln\frac{W_o(S|x)-\delta}{W_o(S|x')}\right)^+,
\end{equation}

\noindent where $\beta^+=\max\{0,\beta\}$. Clearly, for a given $\delta$, $\epsilon^*(\delta,\rho,P_X)$ is 
expected to be nondecreasing in $\rho$. This is verified next by a simple calculation when $P_X$ 
is the uniform distribution on $\cX$. Then, it suffices to fix any $x\neq x'$ in $\cX$ with 
$f(x)\neq f(x')$, and limit the maximum in~\eqref{eq:eps_diff_privacy} to be over only $S\subseteq \cX$. 
By~\eqref{eq:optchannel1},
\begin{equation}
\label{eq:diff_privacy_optchannel1}
W_o(S|x)=\max\left\{\frac{1}{k},\rho\right\}\mathbbm{1}(f(x)\in S) + 
\frac{1-\max\left\{\frac{1}{k},\rho\right\}}{k-1}\left(|S|-\mathbbm{1}(f(x)\in S)\right),
\end{equation}

\noindent and so the maximum in~\eqref{eq:eps_diff_privacy} is attained by $S\subseteq \cX$ 
such that $f(x)\in S$ and $f(x')\notin S$. Hence, from~\eqref{eq:eps_diff_privacy},
\begin{equation}
\label{eq:diff_privacy_expression}
\epsilon^*(\delta,\rho)=\max_{1\leq |S|\leq k-1}\left[\ln\left(\frac{\max\left\{\frac{1}{k},\rho\right\} + 
\left(|S|-1\right)\frac{\left(1-\max\left\{\frac{1}{k},\rho\right\}\right)}{k-1}-\delta}
{\frac{|S|\left(1-\max\left\{\frac{1}{k},\rho\right\}\right)}{k-1}}\right)\right]^+.
\end{equation}

\noindent When $\rho\leq 1/k$, a straightforward calculation shows that $\epsilon^*(\delta,\rho)=0$ 
(which also follows trivially from $W_o$ having all-identical rows). For the case 
$1/k<\rho< 1$,~\eqref{eq:diff_privacy_expression} reduces to
\begin{equation*}
\epsilon^*(\delta,\rho)=\max_{1\leq |S|\leq k-1}\left[\ln\left(1+\frac{\rho-\frac{1-\rho}{k-1}-\delta}
{|S|\frac{1-\rho}{k-1}} \right)\right]^+.
\end{equation*}

\noindent If $\rho- (1-\rho)/(k-1) \geq\delta$, i.e., $\rho\geq 1/k+\delta
(k-1)/k$, then 
\[\epsilon^*(\delta,\rho)=\ln\left(\frac{(k-1)(\rho-\delta)}{1-\rho}\right),\]
noting that $\rho\geq\delta$; and if $1/k<\rho<1/k+\delta(k-1)/k$, $\epsilon^*(\delta,\rho)=0$. 
Summarizing, for a given $\delta\in[0,1]$, $(\epsilon,\delta)$-differential privacy is attainable 
by $W_o:\cX\rightarrow\cZ$ in~\eqref{eq:optchannel1} for all $\epsilon\geq\epsilon^*(\delta,\rho)$, where
\[
\epsilon^*(\delta,\rho)=
\begin{cases}
0,&0\leq\rho<\frac{1}{k}+\frac{\delta(k-1)}{k}\\
\ln\left(\frac{(k-1)(\rho-\delta)}{1-\rho}\right),\hspace{4mm}&\frac{1}{k}+\frac{\delta(k-1)}{k}\leq\rho< 1.
\end{cases}
\]
\par Finally, for the i.i.d. add-noise $\rho$-QRs with $V_1:\cZ\rightarrow\cZ$ and $V_2:\cZ\rightarrow\cZ$ 
of Section~\ref{sec:rhoprivn}, $(\epsilon,\delta)$-differential privacy can be defined as above for associated 
$W_1:\cX\rightarrow\cZ$ and $W_2:\cX\rightarrow\cZ$ obtained by means of Lemma~\ref{lem:addnoise}, with resulting 
privacies $\pi_{\rho}\left(W_1^n\right)=\pi_{\rho}\left(V_1^n\right)$ and $\pi_{\rho}\left(W_2^n\right)
=\pi_{\rho}\left(V_2^n\right)$ by Lemma~\ref{lem:privacy-VW-equiv}(ii). It is easily verified now that 
$(\epsilon,\delta)$-differential privacy is attainable \textit{only} for $\delta=1$ and 
$\epsilon^*\left(1,\rho,P_X\right)=0$, for any $P_X$.
\fi



\appendices

\section{Proof of Achievability in Proposition~\ref{prop:predicate}}
\label{app:predicate_achiev}

\noindent We have that $1 - \pi'_{\rho}\left(W'_o\right)$ equals
\begin{align}
P\big(g'_{MAP(W'_o)}\left(Z\right)= Y\big) &= \sum\limits_{i\in\cZ} \max_{j\in\cY} P\left(Z=i,Y=j\right)\nonumber\\
&= \sum\limits_{i\in\cZ} \max_{j\in\cY} \sum\limits_{x\in h^{-1}(j)} P_X(x)W'_o(i|x).\label{eq:app_predicate_proof_priv}
\end{align}

\noindent When $\rho'_c=1$, we get in~\eqref{eq:app_predicate_proof_priv}, upon using~\eqref{eq:predicate_channel_trivial}, 
that 
\begin{equation}
\label{eq:pred_app_proof_rhoc_1}
P\big(g'_{MAP(W'_o)}\left(Z\right)= Y\big) = \sum\limits_{i\in\cZ} \max_{j\in\cY} 
\sum\limits_{x\in h^{-1}(j)\cap f^{-1}(i)} P_X(x)=\sum\limits_{i\in\cZ}P_X\left(i,j_i^*\right).
\end{equation}

\noindent When $\rho'_c<1$, for each $i\in\cZ$ the 
summand in~\eqref{eq:app_predicate_proof_priv}, upon using~\eqref{eq:predicate_channel}, is

\begin{flalign}
&\max_{j\in\cY}\left[\sum\limits_{x\in h^{-1}(j)\cap f^{-1}(i)} P_X(x)W'_o(i|x) + 
\sum\limits_{x\in h^{-1}(j)\setminus f^{-1}(i)} P_X(x)W'_o(i|x)\right]&\nonumber\\
&=\max_{j\in\cY}\left[\sum\limits_{x\in h^{-1}(j)\cap f^{-1}(i)} P_X(x)\left\{\max\{\rho'_c,\rho\}+
\left(1-\max\{\rho'_c,\rho\}\right) \frac{P_X\left(i,j_i^*\right)-P_X(i,j)}{\sum\limits_{l\in\cZ}
P_X\left(l,j_l^*\right)-P_X\left(h^{-1}(j)\right)}\right\}\right. \nonumber&\\
   & \left.  + \sum\limits_{x\in h^{-1}(j)\setminus f^{-1}(i)} P_X(x)\left\{\left(1-\max\{\rho'_c,\rho\}\right)
\frac{P_X\left(i,j_i^*\right)-P_X(i,j)}{\sum\limits_{l\in\cZ}P_X\left(l,j_l^*\right)-P_X\left(h^{-1}(j)\right)} 
\right\}\right] \nonumber&\\
&=\max_{j\in\cY} \left[
\max\{\rho'_c,\rho\}P_X(i,j) + \left(1-\max\{\rho'_c,\rho\}\right)\frac{P_X\left(i,j_i^*\right)-P_X(i,j)}
{\sum\limits_{l\in\cZ}P_X\left(l,j_l^*\right)-P_X\left(h^{-1}(j)\right)}P_X\left(h^{-1}(j)\right)\right].
\label{eq:pred_achiv_proof}&
\end{flalign}

\noindent It suffices to show that the right-side of~\eqref{eq:pred_achiv_proof} is bounded above by 
$\max\{\rho'_c,\rho\}P_X\left(i,j_i^*\right)$ for each $i\in\cZ$; this is done below. Then, 
in fact, the right-side of~\eqref{eq:pred_achiv_proof} equals $\max\{\rho'_c,\rho\}P_X\left(i,j_i^*\right)$ as seen 
by setting $j=j_i^*$ in the term within $[\cdots]$. 

\par First consider the case when $\max\{\rho'_c,\rho\}<1$. It is seen from~\eqref{eq:predicate_rhoc} 
that for each $j\in\cY$,
\begin{equation}
 \label{eq:pred_achiv_proof_rhoc}
 \frac{\max\{\rho'_c,\rho\}}{1-\max\{\rho'_c,\rho\}} \geq 
 \frac{\rho'_c}{1-\rho'_c}= 
  \frac{P_X\left(h^{-1}(j^*)\right)}
 {\sum\limits_{l\in\cZ}P_X\left(l,j_l^*\right)-P_X\left(h^{-1}(j^*)\right)}
 \geq \frac{P_X\left(h^{-1}(j)\right)}
 {\sum\limits_{l\in\cZ}P_X\left(l,j_l^*\right)-P_X\left(h^{-1}(j)\right)}.
\end{equation}

\noindent Using~\eqref{eq:pred_achiv_proof_rhoc} in~\eqref{eq:pred_achiv_proof}, and since 
$P_X\left(h^{-1}(j)\right)>0$, $j\in\cY$, we get that the right-side of~\eqref{eq:pred_achiv_proof} 
is bounded above by $\max\{\rho'_c,\rho\}P_X\left(i,j_i^*\right)$. Also, this is true trivially 
when $\max\{\rho'_c,\rho\}=1$. Hence, we get that for each $i\in\cZ$, the right-side of~\eqref{eq:pred_achiv_proof} 
equals $\max\{\rho'_c,\rho\}P_X\left(i,j_i^*\right)$. This, combined 
with~\eqref{eq:app_predicate_proof_priv}-\eqref{eq:pred_achiv_proof}, yields
\[
\pi'_{\rho}\left(W'_o\right) = 1-\max\{\rho'_c,\rho\}\sum\limits_{i\in\cZ}P_X\left(i,j_i^*\right).
\]
\qeed


\section{Proof of Lemma~\ref{lem:binom-bds}}

\label{app:lemm_bds}
\begin{enumerate}[(i)]
\item  For each $0\leq\rho\leq 1,$
\[P\left(\text{\normalfont{Bin}}
(n,\rho)\leq\left\lfloor\frac{n}{2}\right\rfloor\right)=
\sum\limits_{t=0}^{\left\lfloor\frac{n}{2}\right\rfloor}P\left(T_{\text{
\normalfont{Ber}}\left(\frac{t}{n}\right)}\right)\]

where $T_{\text{
\normalfont{Ber}}\left(\frac{t}{n}\right)}$ denotes the set of all $n$-length binary sequences of 
``type'' Ber$\left(\frac{t}{n}\right)$, i.e., with $t$ $1$s (and $(n-t)$ $0$s), so that 

\begin{equation}
\label{eq:app-1}
\max_{0\leq t\leq\left\lfloor\frac{n}{2}\right\rfloor}
P\left(T_{\text{\normalfont{Ber}}\left(\frac{t}{n}\right)}\right)\leq P\left(\text{
\normalfont{Bin}}(n,\rho)\leq\left\lfloor\frac{n}{2}\right\rfloor\right)\leq
\left(\left\lfloor\frac{n}{2}\right\rfloor+1\right)\max_{0\leq t\leq\left\lfloor\frac{n}{2}\right\rfloor}
P\left(T_{\text{\normalfont{Ber}}\left(\frac{t}{n}\right)}\right).
\end{equation}

Using well-known bounds for the probability of all $n$-length sequences of a given type 
(cf. {\cite[Lemma $2.6$]{CsiszarKorner11}}), for each $0\leq\rho\leq 1$, and noting that the 
number of types for binary sequences of length $n$ equals $n+1$, 
\begin{equation}
\label{eq:app-2}
\frac{1}{n+1}
\exp\left[-nD\left(\text{\normalfont{Ber}}\left(\frac{t}{n}\right)
\Big|\Big|\text{\normalfont{Ber}}\left(\rho\right)\right)\right]
\leq P\left(T_{\text{\normalfont{Ber}}\left(\frac{t}{n}\right)}\right)\leq
\exp\left[-nD\left(\text{\normalfont{Ber}}\left(\frac{t}{n}\right)
\Big|\Big|\text{\normalfont{Ber}}\left(\rho\right)\right)\right]
\end{equation} 

and noting that for $0.5\leq\rho\leq 1,$
\begin{equation}
\label{eq:app-3}
\min_{0\leq t\leq\left\lfloor\frac{n}{2}\right\rfloor}
D\left(\text{\normalfont{Ber}}\left(\frac{t}{n}\right)\Big |\Big|\text{\normalfont{Ber}}
\left(\rho\right)\right)=D\left(\text{\normalfont{Ber}}\left(\frac{1}{n}\left\lfloor\frac{n}{2}
\right\rfloor\right)\Big |\Big|\text{\normalfont{Ber}}\left(\rho\right)\right)
\end{equation}

we have, by~\eqref{eq:app-2} and~\eqref{eq:app-3}, from~\eqref{eq:app-1} that
\begin{multline*}
\frac{1}{n+1}
\exp\left[-nD\left(\text{\normalfont{Ber}}\left(\frac{1}{n}\left\lfloor\frac{n}{2}
\right\rfloor\right)\Big|\Big|\text{\normalfont{Ber}}\left(\rho\right)\right)\right]\leq 
P\left(\text{\normalfont{Bin}}(n,\rho)\leq\left\lfloor\frac{n}{2}\right\rfloor\right)\\\leq
\left(\left\lfloor\frac{n}{2}\right\rfloor+1\right)\exp\left[-nD\left(\text{\normalfont{Ber}}
\left(\frac{1}{n}\left\lfloor\frac{n}{2}\right\rfloor\right)\Big|\Big|\text{\normalfont{Ber}}
\left(\rho\right)\right)\right].
\end{multline*}

\item We have that
\begin{align*}
P\left(\text{\normalfont{Bin}}(n,\rho)\geq\left\lfloor\frac{n}{2}\right\rfloor+1\right)
&=\sum\limits_{t=\left\lfloor\frac{n}{2}\right\rfloor+1}^{n}{n \choose t}\rho^t\left(1-\rho\right)^{n-t}\\
&=\left(1-\rho\right)^n\sum\limits_{t=\left\lfloor\frac{n}{2}\right\rfloor+1}^{n}{n \choose t}
\left(\frac{\rho}{1-\rho}\right)^t\\
&\leq \left(1-\rho\right)^n\left(\frac{\rho}{1-\rho}\right)^{\left\lfloor\frac{n}{2}\right\rfloor+1}
\sum\limits_{t=\left\lfloor\frac{n}{2}\right\rfloor+1}^{n}{n \choose t},\text{  since $0\leq\rho\leq 0.5$}\\
&\leq \left(1-\rho\right)^n\left(\frac{\rho}{1-\rho}\right)^{\left\lfloor\frac{n}{2}\right\rfloor+1}2^{n-1}\\
&\leq \left(1-\rho\right)^n\left(\frac{\rho}{1-\rho}\right)^{\frac{n-1}{2}+1}2^{n-1}\\
&=\rho\left(2\sqrt{\rho(1-\rho)}\right)^{n-1}\\
&\leq\rho,\hspace{4mm}\text{since $2\sqrt{\rho(1-\rho)}\leq 1$ for $0\leq\rho\leq 1.$}
\end{align*}

\noindent The assertion follows. 
\end{enumerate}\qeed



\section{Proof of~\eqref{eq:func_estimate_lb}}
\label{app:funct_estimate_lb}

\noindent Since $\rho$ and $\max_{i\in\cZ}\hspace{1mm}P\left(f(X)=i\right)$ are obvious lower bounds 
for the left-side of~\eqref{eq:func_estimate_lb}, it suffices to show that
\begin{equation}
 \label{eq:func_estimate_lb_pf1}
 P\big(h_{\text{MAP}}\left(F_1(X),\ldots,F_n(X)\right)= f(X)\big) \geq 
 P\left(\text{\normalfont{Bin}}(n,\rho)\geq\left\lfloor\frac{n}{2}\right\rfloor+1\right).
\end{equation}

\noindent The proof bears a resemblance to that of Theorem~\ref{thm:rhoprivn-converse} above and so we shall refer 
to pertinent details therein. We have
\begin{multline}
 \label{eq:func_estimate_lb_pf2}
 P\big(h_{\text{MAP}}\left(F_1(X),\ldots,F_n(X)\right)= f(X)\big)\\ = 
 \sum\limits_{\left(i_1,\ldots,i_n\right)\in\cZ^n}\max_{
j\in\cZ}\hspace{1.5mm}P\left(f(X)=j\right)P\left(F_1(X)=i_1,\ldots,F_n(X)=i_n|f(X)=j\right).
\end{multline}

\noindent Since 
\begin{align*}
 \hspace{-2mm}&P\left(F_1(X)=i_1,\ldots,F_n(X)=i_n|f(X)=j\right)\\&\hspace{12mm} 
 =\sum\limits_{x\in\cX} P\left(F_1(X)=i_1,\ldots,F_n(X)=i_n|f(X)=j,X=x\right)P\left(X=x|f(X)=j\right)\\
 &\hspace{12mm}=\sum\limits_{x\in f^{-1}(j)} \prod\limits_{t=1}^n W_t\left(i_t|x\right)
 \frac{P_X(x)}{P\left(f(X)=j\right)},
\end{align*}

\noindent we get in~\eqref{eq:func_estimate_lb_pf2} with $\cA_l(i)$ in~\eqref{eq:sign_terms} that
\begin{align*}
 P\big(h_{\text{MAP}}\left(F_1(X),\ldots,F_n(X)\right)= f(X)\big)
 &=\sum\limits_{\left(i_1,\ldots,i_n\right)\in\cZ^n}\max_{
j\in\cZ}\hspace{1.5mm}\left(\sum\limits_{x\in f^{-1}(j)}P_X(x) 
\prod\limits_{t=1}^n W_t\left(i_t|x\right)\right)\\
&\geq \sum\limits_{i\in\cZ}\sum\limits_{l=\left\lfloor\frac{n}{2}
\right\rfloor+1}^{n}\sum\limits_{\left(i_1,\ldots,i_n\right)\in\cA_l(i)}\max_{
j\in\cZ}\hspace{1.5mm}\left(\sum\limits_{x\in f^{-1}(j)}P_X(x) 
\prod\limits_{t=1}^n W_t\left(i_t|x\right)\right)\\
&\geq\sum\limits_{i\in\cZ}\sum\limits_{x\in f^{-1}(i)}P_X(x)\left(\sum\limits_{l=\left\lfloor\frac{n}{2}
\right\rfloor+1}^{n}\sum\limits_{\left(i_1,\ldots,i_n\right)\in\cA_l(i)}
\prod\limits_{t=1}^n W_t\left(i_t|x\right)\right).
\end{align*}

\noindent Mimicking~\eqref{eq:n-privacyub3}-\eqref{eq:n-privacyub5a}, observe that the sum above 
within $(\cdot)$ is bounded below by $P\left(\text{\normalfont{Bin}}(n,\rho)\geq\left\lfloor
\frac{n}{2}\right\rfloor+1\right)$. Clearly, ~\eqref{eq:func_estimate_lb_pf1} follows.\qeed



\section{Proof of Proposition~\ref{prop:Vo-chernoff}($\text{\lowercase{\normalfont{i}}}$)} 
\label{app:prop_chernoff}
\setcounter{theorem}{0}
\renewcommand{\thelemma}{\Alph{section}\arabic{lemma}}
\counterwithin{theorem}{section}

\noindent The following two lemmas are pertinent. Recall from~\eqref{eq:prob} that 
$x^*=\arg\hspace{0.1cm}\max_{x\in\cX}\hspace{1.5mm}P_X\left(x\right)$ is in 
$f^{-1}\left(i^*\right)$ for some (fixed) $i^*\in\cZ$.

\begin{lemma}
\label{lem:appendix1}
For $V_o:\cZ\rightarrow\cZ$ in~\eqref{eq:optchannel2},
\begin{enumerate}[(i)]
\item when $\rho_c<\rho\leq 1$, no two rows can be identical;
\item when $0\leq \rho\leq\rho_c$, if the rows $V_o(\cdot|j)$ and $V_o(\cdot|j')$, 
$j\neq j'$, are identical, then each coincides with the row $V_o(\cdot|i^*)$, in which 
case $P_X\big(x_j^*\big)=P_X\big(x_{j'}^*\big)=P_X\left(x^*\right)$. Furthermore, the number of identical 
rows of $V_o$ cannot exceed $\left\lfloor\frac{1}{\rho_c}\right\rfloor.$
\end{enumerate}
\end{lemma}

\vspace{0.2cm}

\noindent\textit{Proof}:\hspace{0.2cm}With $0\leq\rho\leq 1$, if the rows of $V_o:
\cZ\rightarrow\cZ$ corresponding to $j\neq j'$ in $\cZ$ are identical, then
\[
V(i|j)=V(i|j'),\hspace{4mm}i\in\cZ\setminus\{j,j'\}\\
\]

i.e.,
\[
\left(1-\max\{\rho_c,\rho\}\right)\frac{P_X\left(x_i^*\right)}
{\sum\limits_{l\neq j}P_X\left(x_l^*\right)}=\left(1-\max\{\rho_c,\rho\}\right)
\frac{P_X\left(x_i^*\right)}{\sum\limits_{l'\neq j'}P_X\left(x_{l'}^*\right)},
\hspace{4mm}i\in\cZ\setminus\{j,j'\}
\]

whence
\begin{equation}
\label{eq:lemmappendix1}
P_X\left(x_j^*\right)=P_X\left(x_{j'}^*\right);
\end{equation}

and furthermore
\[
V(i|j)=V(i|j'),\hspace{4mm}i\in\{j,j'\} 
\]

which, using~\eqref{eq:lemmappendix1}, gives straightforwardly that
\begin{equation}
\label{eq:lemmappendix2}
\max\{\rho_c,\rho\} = \frac{P_X\left(x_i^*\right)}{\sum\limits_{l\in\cZ}P_X\left(x_l^*\right)},
\hspace{4mm}i\in\{j,j'\}.
\end{equation}

\begin{enumerate}[(i)]
\item When $\rho>\rho_c$, recalling~\eqref{eq:rhoc}
\[\frac{P_X\left(x_i^*\right)}{\sum\limits_{l\in\cZ}P_X\left(x_l^*\right)}\leq \frac{P_X\left(x^*\right)}
{\sum\limits_{l\in\cZ}P_X\left(x_l^*\right)} =\rho_c<\max\{\rho_c,\rho\}\]
which violates~\eqref{eq:lemmappendix2} for $i\in\{j,j'\}$, so that no two 
rows of $V_o:\cZ\rightarrow\cZ$ can be identical.
\item When $0\leq\rho\leq\rho_c$, suppose that the rows $V_o(\cdot|j)$ and $V_o(\cdot|j')$ 
are identical for some $j\neq j'$. Then~\eqref{eq:lemmappendix2} holds which, upon 
recalling~\eqref{eq:rhoc}, is tantamount to
\begin{equation}
\label{eq:lemmappendix3}
P_X\big(x_j^*\big)=P_X\big(x_{j'}^*\big)=P_X(x^*)=P_X(x_{i^*}^*). 
\end{equation}

\noindent To show for $j\neq i^*$ that $V_o(i|j)=V_o(i|i^*)$, $i\in\cZ$, consider first 
$i\in\{j,i^*\}$. Then, using~\eqref{eq:lemmappendix3},
\[V_o(j|j)=\rho_c,\hspace{4mm} V_o(i^*|j)=\left(1-\rho_c\right)\frac{P_X\left(x_{i^*}^*\right)}
{\sum\limits_{l\neq j}P_X\left(x_{l}^*\right)}=\left(1-\rho_c\right)\frac{P_X\left(x^*\right)}
{\sum\limits_{l\in\cZ}P_X\left(x_{l}^*\right)-P_X\big(x_{j}^*\big)}=\rho_c,
\]
and similarly,
\[
V_o(j|i^*)=\left(1-\rho_c\right)\frac{P_X\big(x_{j}^*\big)}
{\sum\limits_{l\in\cZ}P_X\left(x_{l}^*\right)-P_X\left(x_{i^*}^*\right)}=\rho_c,\hspace{4mm}V_o(i^*|i^*)=\rho_c.
\]
And for $i\in\cZ\setminus\{j,i^*\}$,
\[V_o(i|j)=\left(1-\rho_c\right)\frac{P_X\left(x_{i}^*\right)}
{\sum\limits_{l\in\cZ}P_X\left(x_{l}^*\right)-P_X\big(x_{j}^*\big)}=
\frac{P_X\left(x_{i}^*\right)}{\sum\limits_{l\in\cZ}P_X\left(x_{l}^*\right)}=V_o(i|i^*).\]
\par Lastly, if the number of identical rows of $V_o:\cZ\rightarrow\cZ$ is $\alpha$, then 
$\alpha P_X\left(x^*\right)\leq \sum\limits_{l\in\cZ}P_X\left(x_l^*\right)$, 
whence $\alpha\leq \left\lfloor\frac{1}{\rho_c}\right\rfloor.$ 
\end{enumerate}

\qeed

\vspace{0.2cm}
\par For $S\subseteq\cZ,$ let 
\begin{equation}
\label{eq:app_probdefn}
j_S=\arg\hspace{0.1cm}\max_{l\in 
S}\hspace{1.5mm}P_X\left(x_l^*\right),\hspace{5mm} 
R_S=\left(\cZ\setminus S\right) \cup \{j_{S}\},
\end{equation}

\noindent where $j_S$ and $R_S$ need not be unique. Let $\tilde{f}_{R_S}(X)$ be a 
$R_S$-valued rv with pmf
\begin{equation}
\label{eq:applemmaprobability}
P\left(\tilde{f}_{R_S}(X)=i\right)=\frac{P_X\left(x_i^*\right)}{\sum\limits_{l\in R_S}
P_X\left(x_l^*\right)},\hspace{4mm}i\in R_S.
\end{equation}

\noindent Consider the stochastic matrix $V_{R_S}:R_S\rightarrow\cZ$ given by 
\begin{equation}
\label{eq:app_lemm_matrix}
V_{R_S}=\{V(i|j),i\in\cZ,j\in R_S\}
\end{equation}

\noindent and let $\left\{\tilde{Z}_t^{R_S}\right\}_{t=1}^n$ be conditionally mutually independent 
$\cZ$-valued rvs conditioned on $\tilde{f}_{R_S}(X),$ with 
\begin{equation} 
\label{eq:applemmaaddnoise}
P_{\tilde{Z}_t^{R_S}|\tilde{f}_{R_S}(X)}=V_{R_S},\hspace{4mm}t=1,\ldots,n.
\end{equation}

\noindent Let $C\left(V_{R_S}\right)$ be the Chernoff radius restricted to $R_S$, i.e., with the 
minimum in~\eqref{eq:chern-rad} being instead over all $j\neq j'$ in $R_S$.

\begin{lemma}
\label{lem:appendix2}
For $0\leq\rho\leq 1,$ consider add-noise $\rho$-QRs $\left\{F_t(X)\right\}_{t=1}^n$
 with~\eqref{eq:addnoisen} holding for every $n\geq 1,$ where $V_t=V,$ $t\geq 1.$ 
If $V:\cZ\rightarrow\cZ$ has identical rows $\left\{V(\cdot|j),j\in S\right\}$, then 
\[\pi_{\rho}\left(V^n\right) - \left(1-\sum\limits_{i\in R_S}P_X\left(x_i^*\right)\right)\doteq 
\exp\left[-nC\left(V_{R_S}\right)\right]\]

\noindent for $R_S$ and $V_{R_S}$ in~\eqref{eq:app_probdefn} 
and~\eqref{eq:app_lemm_matrix}, respectively.
\end{lemma} 

\noindent \textit{Remark}: If the rows of $V_{R_S}:R_S\rightarrow\cZ$ are 
distinct in Lemma~\ref{lem:appendix2}, then $C\left(V_{R_S}\right)>0$. If the rows of 
$V:\cZ\rightarrow\cZ$ are distinct, then $S=\phi$, $R_S=\cZ$ and $V_{R_S}=V$.\\

\noindent\textit{Proof}: 
\begin{align}
 P\left(g_{\text{MAP$\left(V^n\right)$}}\left(Z_1,\ldots,Z_n\right)=X\right)
 &=\sum\limits_{\left(i_1,\ldots,i_n\right)\in\cZ^n}\max_{
x\in\cX}\hspace{1.5mm}P_X(x)\prod\limits_{t=1}^{n}V\left(i_t|f(x)\right)\nonumber\\
&=\sum\limits_{\left(i_1,\ldots,i_n\right)\in\cZ^n}\max_{\substack{x\in\cup f^{-1}(j) \\ j\in\cZ}}
\hspace{1.5mm}P_X(x)\prod\limits_{t=1}^{n}V\left(i_t|f(x)\right)\nonumber\\
&=\sum\limits_{\left(i_1,\ldots,i_n\right)\in\cZ^n}\max_{
j\in\cZ}\hspace{1.5mm}P_X\left(x_j^*\right)\prod\limits_{t=1}^{n}
V\left(i_t|j\right)\hspace{4mm}\nonumber\\
&=\sum\limits_{\left(i_1,\ldots,i_n\right)\in\cZ^n}\max_{
j\in(\cZ\setminus S)\cup S}\hspace{1.5mm}P_X\left(x_j^*\right)
\prod\limits_{t=1}^{n}V\left(i_t|j\right)
\hspace{4mm}\nonumber\\
&=\sum\limits_{\left(i_1,\ldots,i_n\right)\in\cZ^n}\max_{
j\in R_S}\hspace{1.5mm}P_X(x_j^*)\prod\limits_{t=1}^{n}
V\left(i_t|j\right)\label{eq:privacy-applemma-equiv}\\
&=\left(\sum\limits_{i\in R_S}P_X\left(x_i^*\right)\right)
\sum\limits_{\left(i_1,\ldots,i_n\right)\in\cZ^n}
\max_{j\in R_S}\hspace{1.5mm}\frac{P_X\big(x_j^*\big)}
{\sum\limits_{i\in R_S}P_X\left(x_i^*\right)}
\prod\limits_{t=1}^{n}V\left(i_t|j\right)\nonumber\\
&=\left(\sum\limits_{i\in R_S}P_X\left(x_i^*\right)\right) 
P\left(g_{\text{MAP$\left(V_{R_S}^n\right)$}}
\left(\tilde{Z}^{R_S}_1,\ldots,\tilde{Z}^{R_S}_n\right)=
\tilde{f}_{R_S}(X)\right)\label{eq:privacy-applemma-equiv-proof}
\end{align}

\noindent where~\eqref{eq:privacy-applemma-equiv} is by the identicality of the rows  
$\{V\left(\cdot|j\right),\text{ }j\in S\}$, and $\tilde{f}_{R_S}(X)$ and 
$\{\tilde{Z}^{R_S}_t\}_{t=1}^n$ are as in~\eqref{eq:applemmaprobability} 
and~\eqref{eq:applemmaaddnoise}, respectively. The assertion 
follows by applying {\cite[Theorem $2$]{KanayaHan95}} 
to~\eqref{eq:privacy-applemma-equiv-proof}.\qeed\\

\par Turning to the proof of Proposition~\ref{prop:Vo-chernoff}(i), 
first observe by Theorem~\ref{thm:rhoprivn-achiev2} that for $0\leq\rho\leq 0.5$ and \textit{every} $n\geq 1$,

\begin{align}
\pi_{\rho}\left(V_2^n\right) - \left(1-\sum\limits_{i\in\cZ}P_X\left(x_i^*\right)\right)&\geq  
\sum\limits_{i\in\cZ}P_X\left(x_i^*\right) - \max\left\{P_X(x^*),\sum\limits_{i=0}^{\left\lfloor\frac{k}
{\left\lfloor
\frac{1}{\rho}\right\rfloor}\right\rfloor-\mathbbm{1}\left(k\text{ \normalfont{mod}}\left\lfloor\frac{1}
{\rho}\right\rfloor=0\right)}
       P_X\left(x_{i\left\lfloor\frac{1}{\rho}\right\rfloor}^*\right)\right\} \nonumber \\ 
&>0. \label{eq:applemma-positivity}
\end{align}

We consider two cases: $\rho>\rho_c$ and $0\leq\rho\leq\rho_c$. 
\par When $\rho>\rho_c$, by Lemma~\ref{lem:appendix1}(i), all the rows of 
$V_o:\cZ\rightarrow\cZ$ are distinct so that $C(V_o)>0$. Then, by~\eqref{eq:prop_chernoff_compare1},
\begin{equation*}
\lim_n\hspace{1.5mm} \pi_{\rho}\left(V_o^n\right) - \left(1-\sum\limits_{i\in\cZ}P_X\left(x_i^*\right)\right)=0
\end{equation*}

\noindent which upon comparison with~\eqref{eq:applemma-positivity}, 
yields~\eqref{eq:prop-chernoff0} in this case.

\par In the case $0\leq\rho\leq\rho_c$, $V_o:\cZ\rightarrow\cZ$ can contain identical rows. 
By Lemma~\ref{lem:appendix1}(ii) and upon invoking assumption~\eqref{eq:prob-assumption} 
without loss of generality, the identical rows must be those corresponding to $\{0,1,\ldots,a-1\}$ 
(with the remaining rows being all distinct), where $a\leq\left\lfloor\frac{1}{\rho_c}\right\rfloor$ is 
the number of identical rows. By applying Lemma~\ref{lem:appendix2}, with $S=\{0,1,\ldots,a-1\},$ $j_S={0}$ 
and observing that $C\left(\left(V_o\right)_{R_S}\right)>0,$ we get
\begin{align*}
\lim_n \hspace{1.5mm} \pi_{\rho}\left(V_o^n\right) 
&= 1-\sum\limits_{i\in R_S}P_X\left(x_i^*\right)\\
&=\begin{cases}
 1-P_X\left(x^*\right),\hspace{4mm} &a=k
 \Leftrightarrow\left\lfloor\frac{1}{\rho_c}\right\rfloor=k\\[7.5pt]
 1-\sum\limits_{i\in \{0,a,a+1,\ldots,k-1\}}P_X\left(x_i^*\right),
 \hspace{4mm}&a<k\Leftrightarrow\left\lfloor\frac{1}{\rho_c}\right\rfloor<k
\end{cases}\\
&\leq\begin{cases}
1-P_X\left(x^*\right),\hspace{4mm}&\left\lfloor\frac{1}{\rho_c}\right\rfloor=k\\[7.5pt]
 1-\sum\limits_{i\in\left\{0,\left\lfloor\frac{1}{\rho_c}\right\rfloor,\left\lfloor\frac{1}{\rho_c}
 \right\rfloor+1,\ldots,k-1\right\}}
 P_X\left(x_i^*\right),\hspace{4mm} &\left\lfloor\frac{1}{\rho_c}\right\rfloor<k
 \end{cases}\\
&\leq \begin{cases}
 1 - P_X\left(x^*\right), &\rho_c=\frac{1}{k}\\
 1 - \sum\limits_{i=0}^{\left\lfloor\frac{k}{\left\lfloor\frac{1}{\rho_c}\right\rfloor}\right\rfloor}
       P_X\left(x_{i\left\lfloor\frac{1}{\rho_c}\right\rfloor}^*\right),
       &k\text{ \normalfont{mod}}\left\lfloor\frac{1}
{\rho_c}\right\rfloor\neq 0,\hspace{4mm}  \frac{1}{k} < \rho_c\leq 0.5\\
 1 - \sum\limits_{i=0}^{\left\lfloor\frac{k}{\left\lfloor\frac{1}{\rho_c}\right\rfloor}\right\rfloor-1}
       P_X\left(x_{i\left\lfloor\frac{1}{\rho_c}\right\rfloor}^*\right),
       &k\text{  \normalfont{mod}}\left\lfloor\frac{1}
{\rho_c}\right\rfloor=0,\hspace{4mm} \frac{1}{k} < \rho_c\leq 0.5                                           
 \end{cases} \\
&= \pi_{\rho_c}\left(V_2^n\right)\\
&\leq \pi_{\rho}\left(V_2^n\right),\hspace{4mm}0\leq\rho\leq\rho_c,
\end{align*} 

\noindent which, upon recalling by Theorem~\ref{thm:rhoprivn-achiev2} that $\pi_{\rho}\left(V_2^n\right)$
is the same for all $n$, establishes~\eqref{eq:prop-chernoff0}. \qeed
 


\section*{Acknowledgement}
The authors are indebted to: Himanshu Tyagi and Shun Watanabe for educating us in private 
function computation, helping formulate the models in this paper, and for numerous beneficial 
discussions that informed our approach; Adam Smith for raising the question of predicate 
privacy~\cite{Smith18}; and all three anonymous reviewers for examining our work under a microscope 
and suggesting several improvements.



\end{document}